\pdfoutput=1
\documentclass[a4paper,10pt,amsmath,amssymb,nofootinbib,onecolumn,floats,longbibliography,aps,prd]{revtex4-2}

\usepackage[british]{babel}
\usepackage{txfonts}
\usepackage{mathrsfs,bm}
\usepackage{natbib,textcase}
\usepackage{graphicx,color,soul,comment}
\usepackage{subcaption}

\usepackage[pdftex,pdfusetitle]{hyperref}
\hypersetup{colorlinks=true,linkcolor=red,citecolor=blue,filecolor=black,urlcolor=black,pdfauthor={}}
\usepackage[capitalize]{cleveref}

\usepackage{physics}
\usepackage{array,booktabs} 
\usepackage{multirow} 
\newcolumntype{L}{>{$}p{20mm}<{$}} 
\makeatletter\g@addto@macro\bfseries{\boldmath}\makeatother%

\allowdisplaybreaks

\def\be#1\ee{\begin{align}#1\end{align}}

\newcommand{\ie}{i.e.}

\renewcommand{\dd}{\text{d}}
\newcommand{\e}{\text{e}}

\renewcommand{\geq}{\geqslant}
\renewcommand{\le}{\leqslant}
\renewcommand{\leq}{\leqslant}

\newcommand{\ICC}{\affiliation{Department of Physics, King’s College London, The Strand, London, WC2R 2LS, UK}}

\newcommand{\PI}{\affiliation{Perimeter Institute for Theoretical Physics, 31 Caroline St., Waterloo, ON, N2L 2Y5, Canada}}

\begin{document}
\title{Tidal response of regular black holes}

\author{Chiara Coviello}
\ICC

\author{Vania Vellucci}

\affiliation{Quantum Theory Center (${\hbar}$QTC) \& D-IAS, IMADA at Southern Denmark Univ., Campusvej 55, 5230 Odense M, Denmark}

\author{Luis Lehner}
\PI

\begin{abstract}
    In this work, we investigate the tidal deformability of regular black holes (RBHs). Employing different phenomenological models, we analyze their response to both test fields and gravitational perturbations, interpreting the latter within the framework of Einstein's field equations in the presence of an appropriate exotic matter distribution. Numerical and analytical methods reveal that RBHs exhibit non-trivial tidal responses, influenced by their regularization parameters and exotic matter distributions. The results obtained for test fields and gravitational perturbations are in qualitative agreement. This hints at the possibility that similar conclusions could hold if these spacetimes  were interpreted as solutions of a modified gravitational action. Our findings suggest that RBHs possess distinct, though subtle, tidal signatures, which may serve as observational probes of their internal structure in gravitational wave detections.
\end{abstract}
\maketitle

\section{Introduction}
A spacetime singularity marks a fundamental breakdown in the structure of spacetime. 
In the framework of general relativity, singularities are not just theoretical possibilities; they are unavoidable in specific scenarios, such as in the gravitational collapse or in the past evolution of an expanding universe \cite{PhysRevLett.Penrose, 1970.Penrose.Hawking}. These singularities lead to a loss of predictability and challenge the viability of general relativity in such high-energy scenarios, underscoring the need for a theory that unifies gravity and quantum mechanics.\\
One approach to addressing the singularity problem in black holes suggests that quantum gravity--or, more generally, new physics--modifies their internal structure at a characteristic scale $\ell$, effectively resolving the central singularity. Although the exact form of quantum gravity remains unknown, several phenomenological models have been developed in which the singularity is replaced by a more regular structure. This leads to the concept of regular black holes (RBHs). In spherical symmetry there exist two families of such solutions: in the first, the singularity is replaced by a hypersurface of minimum radius, acting as a wormhole throat hidden within the event horizon, as in the Simpson–Visser model \cite{2019SimpsonVisser}; in the second family, that includes Bardeen-like models \cite{1968Bardeen, 2006Hayward, 2016FanWang}, one introduces an inner horizon that shields a non-singular maximally symmetric core, avoiding the singular breakdown at the center. Notably, it can be demonstrated that these two families of solutions encompass all possible regularized, spherically symmetric black hole spacetimes \cite{2020PhRvDGeodesicallyCompleteBH}. For both types, depending on the value of the regularization parameter $\ell$, configurations with or without horizons can arise. 
These insights hint at a promising avenue for understanding black holes within a quantum framework, potentially reshaping our views of their inner structures and the nature of singularities in extreme gravitational settings. Note that, while the two families of models considered in this work encompass all possible spherically symmetric \textit{regular black holes} (regular objects with a horizon), a vast plethora of alternative \textit{horizonless models} can be constructed. Some examples are gravastars~\cite{MazurMottola2004}, fuzzballs~\cite{Mathur2005,Ikedaetal2021}, semiclassical stars ~\cite{Arrecheaetal2022,Arrecheaetal2023}, frozen stars~\cite{BrusteinMedved2017}, and others~\cite{CardosoPani2017, Raposoetal2018, HoldomRen2017}. Their phenomenology, including their tidal deformability, has been extensively investigated \cite{Kesden:2004qx, Pani:2010em, Cardoso:2016rao, Cardoso:2017cfl, Herdeiro:2020kba,Vellucci:2022hpl,Franzin:2023slm,Berti:2024moe, Arrechea:2024nlp}. \\ 
In this paper, we examine the tidal deformability of RBHs. Tidal effects on self-gravitating objects are typically characterized by a set of coefficients known as tidal Love numbers. It is well established that the static tidal Love numbers of various black holes in four-dimensional spacetimes vanish exactly \cite{2009PhRvDLoveNumberSchwazrschild,Hui:2020xxx,LeTiec:2020bos,Hui:2021vcv,Pani:2015hfa,Sharma:2024hlz}. However, this property does not hold for the aforementioned exotic compact objects \cite{Berti:2024moe,PhysRevD.Love,Pani:2015tga,Herdeiro:2020kba,Chen:2023vet}, nor for black holes in modified theories of gravity \cite{Cardoso:2018ptl,DeLuca:2022tkm,Barura:2024uog,Barbosa:2025uau}, in higher-dimensional spacetimes \cite{Hui:2020xxx,Kol:2011vg,Cardoso:2019vof}, in non-asymptotically flat backgrounds \cite{Franzin:2024cah,Nair:2024mya}, or in non-vacuum environments \cite{Baumann:2018vus,DeLuca:2021ite,Capuano:2024qhv,Katagiri:2023yzm,Cardoso:2019upw,Cannizzaro:2024fpz,Chakraborty:2024gcr}. \\
To investigate the static tidal deformability of RBHs, we study both test fields static perturbations and true gravitational tidal fields. In the test-field approach, we also adopt a model-independent parametrized framework to determine the conditions on the metric functions that allow for a nonzero Love number. For true tidal fields, we must rely on a gravitational theory (a set of field equations) compatible with the considered spacetime solution. Addressing singularities generally requires new physics, whether through quantum gravity, modified gravity or exotic matter within general relativity. In our study of gravitational perturbations, we interpret these spacetimes as solutions of Einstein's field equations in the presence of suitable exotic matter fields--specifically, a magnetic field with a nonlinear lagrangian and, in the case of the Simpson Visser model, an additional phantom scalar field. Unlike the Schwarzschild black hole, these RBHs appear to undergo deformation under tidal fields. The results from test fields and gravitational perturbations, within the context of nonlinear electrodynamics and phantom scalar fields, are in qualitatively agreement, both indicating a non trivial tidal deformation. We take this as a hint that similar results for gravitational perturbations would arise if these spacetimes were interpreted as solutions of a potential quantum, or simply modified, gravitational action \cite{PhysRevD.Platania, PhysRevD.Platania2,collapseRBH,ModifiedGravRBH}.
In this work, we do not scrutinize in detail the implications for gravitational wave observations, given that establishing a direct connection can be nontrivial and ambiguous \cite{2024arXivCanonicalDef}. However, based on the strength of the tidal Love numbers, we estimate (following ~\cite{LNinGW}) that phase differences due to tidal interactions can be as large as a few radians by the time of the plunge towards merger takes place. Such an impact is potentially detectable.  \\
This paper is organized as follows. In Sec. \ref{sec:RBHs}, we review RBHs models. Sec \ref{sec:tidal-deformability} discusses the study of the tidal deformability of compact objects, including ambiguities and different calibrations. In Sec. \ref{sec:numerical-method}, we present the numerical method employed to obtain our results. Sec. \ref{sec:test} presents both the analytical and numerical results obtained through test-field perturbations, and discusses how a nonzero Love number can be obtained in a model-independent way. In Sec. \ref{sec:resultsfullgrav}, we provide the Love number results for both polar and axial gravitational perturbations of the full gravitational system. Additionally, we explore the potential for detecting the different behaviors that distinguish RBHs from singular black holes via gravitational wave observations. Finally, in Sec. \ref{sec:concl}, we summarize our findings and discuss possible directions for future research.\\ Throughout this work, we adopt natural units with $c=G=1$.

\section{Regular Black Hole models}\label{sec:RBHs}
In this paper, we focus on regularized, spherically symmetric, static black hole spacetimes described by the following line element:
\be
ds^2 = -\e^{-2 \phi(r)}f(r)\,\dd t^2+\frac{\dd r^2}{f(r)} + r^2\left(\dd\theta^2 + \sin^2\theta\,\dd\varphi^2\right),
\label{metric}
\ee
where
\be
f(r) = 1 - \frac{2m(r)}{r}.
\ee
The metric coefficients admit an asymptotic expansion in powers of the radial coordinate, introducing corrections to the Schwarzschild metric. These corrections are necessary to bypass Penrose's singularity theorems \cite{PhysRevLett.Penrose, 1970.Penrose.Hawking} and achieve a regular spacetime. In the framework of quantum gravity, they are anticipated, as quantum-gravitational fluctuations modify the effective field equations by introducing higher-order curvature terms, rendering Ricci-flat spacetimes no longer viable solutions \cite{Goroff:1985sz}. We consider the following models: 
\be\label{eq:m(r)-Bardeen}\textit{Bardeen:} \qquad \phi_B(r)=0 \qquad m_B(r)=M \frac{r^3}{(r^2+\ell^2)^{3/2}} \qquad \ell_{extremal}=\frac{4 M}{3 \sqrt{3}},\ee\be \label{eq:m(r)-Hayward}\textit{Hayward:}\qquad \phi_H(r)=0 \qquad m_H(r)=M\frac{r^3}{(r^3+2M\ell^2)} \qquad \ell_{extremal}=\frac{4 M}{3 \sqrt{3}},\ee\be\label{eq:m(r)-FW} \textit{Fan-Wang:}\qquad \phi_{FW}(r)=0\quad m_{FW}(r)=M\frac{r^3}{(r+\ell)^3} \qquad \ell_{extremal}=\frac{8 M}{27},\ee\be\label{eq:m(r)-SV}\textit{Simpson-Visser:}\qquad \phi_{SV}(r)=\frac{1}{2}\log\left(1-\frac{\ell^2}{r^2}\right)\qquad m_{SV}(r)=M\left(1-\frac{\ell^2}{r^2}\right)+\frac{\ell^2}{2r} \qquad \ell_{extremal}=2 M.\ee
All models reduce to the Schwarzschild black hole in the limit $\ell\to0$. The first three models possess two horizons up to a specific value of $\ell$, which we designate as $\ell_\text{extremal}$. At this critical value, they become extremal RBHs, where the two horizons coincide. Beyond $\ell_\text{extremal}$, as $\ell$ continues to increase, the metric describes a compact object without horizons. In contrast, the last listed model features a wormhole throat in the innermost region. Outside this throat, an event horizon exists up to $\ell=2M$, where the wormhole throat matches the size of the horizon, becoming an extremal null throat. Beyond this value of $\ell$, the throat continues to expand, and the metric no longer possesses a horizon. \\
These models can be interpreted within the framework of general relativity as solutions to Einstein's equations, incorporating nontrivial and exotic matter content. For instance, the first three models can be derived in the contest of nonlinear electrodynamics, while the Simpson-Visser spacetime also requires a self-interacting phantom scalar field. The general action is:
\be S=\int d^4 x \sqrt{-g}\left(\frac{1}{16\pi}R-\frac{1}{4\pi}\mathcal{L}(F)+\frac{1}{2}(\partial\Phi)^2-V(\Phi)\right),\ee
where $F=F_{\mu\nu}F^{\mu\nu}/4$ is the electromagnetic field strength with $F_{\mu\nu}=2(\nabla_\mu A_\nu-\nabla_\nu A_\mu)$ and $A_\mu$ the electromagnetic potential; $\Phi$ is the scalar field. It is assumed that the Maxwell field is purely magnetic, with its magnetic charge equal to the regularization parameter $\ell$. The modified Maxwell field equations are:
\be \nabla_\mu (\mathcal{L},_F F^{\nu\mu})=0,\label{eq:modMaxwell}\ee
where $\mathcal{L},_F=\partial\mathcal{L}/\partial F$. The gravitational field equations are:
\be G_{\mu\nu}=2( \mathcal{L},_F F_\mu^\sigma F_{\nu\sigma}-g_{\mu\nu}\mathcal{L})-8\pi \left[\partial_\mu\Phi\partial_\nu\Phi+g_{\mu\nu}\left(-\frac{1}{2}(\partial\Phi)^2+V(\Phi)\right)\right].\label{eq:Einstein}\ee
Finally, the Klein-Gordon equation is derived varying the
action with respect to the scalar field:
\be
\Box \Phi- \frac{\partial V}{\partial \Phi}=0.\label{eq:KG}
\ee
The matter content for each model is:
\be\textit{Bardeen:}\qquad \Phi_{B}=0\qquad \mathcal{L}_{B}=\frac{3M\ell^2}{(r^2+\ell^2)^{5/2}}\qquad V_{B}=0,\ee
\be\textit{Hayward:}\qquad \Phi_{H}=0\qquad \mathcal{L}_{H}=\frac{6M^2\ell^2}{(r^3+2M\ell^2)^2}\qquad V_{H}=0,\ee
\be\textit{Fan-Wang:}\qquad \Phi_{FW}=0\qquad \mathcal{L}_{FW}=\frac{3M\ell}{(r+\ell)^4}\qquad V_{FW}=0,\ee
\be\textit{Simpson-Visser:}\qquad \Phi_{SV}=\frac{1}{\sqrt{4\pi}}\arccot\frac{\sqrt{r^2-\ell^2}}{\ell}\qquad \mathcal{L}_{SV}=\frac{6M}{5}\left(\frac{2F^5}{\ell^2}\right)^{1/4}\qquad V_{SV}=\frac{M\sin^5(\sqrt{4\pi}\Phi)}{10\pi\ell^3}.\ee
Having established the matter content for each model, we now explore how these RBHs respond to an external tidal field. Tidal deformability plays a crucial role in gravitational wave astronomy, as it influences the signals detected by LIGO and Virgo \cite{PhysRevLett.LIGOfirst}. In the next section, we introduce the formalism used to describe tidal deformability and apply it to RBHs.

\section{Tidal deformability}
\label{sec:tidal-deformability}
Compact objects, such as neutron stars, experience deformations when placed in an external tidal field, such as the one generated by a companion in a binary system. These deformations leave measurable imprints on gravitational waves, making tidal deformability a key property in strong-field gravity and observational astrophysics. To quantify this effect, we introduce the gravitational tidal Love numbers, $k_{lm}$, which characterize the body's response to an external perturbation in the far-field regime. In the following, we analyze static linear perturbations of both the RBH metric and the matter fields to determine how these coefficients are defined and computed. Perturbations can be categorized by parity symmetry: odd (axial) and even (polar). By symmetry, polar gravitational perturbations are coupled to axial magnetic and polar scalar perturbations, while axial gravitational perturbations are coupled only to polar magnetic perturbations, since it is not possible to have axial scalar perturbations. In the Regge-Wheeler gauge, after imposing specific relations among the coefficients dictated by the equations and choosing an electromagnetic gauge, the perturbations of the metric $g_{\mu\nu}$, the electromagnetic potential $A_\mu$, and the scalar field $\Phi$, take the following forms:
\be\delta g_{\mu\nu}^{\text{(polar)}}=\text{diag}\left(-f(r) H_0^{lm}(r), -\frac{H_0^{lm}(r)}{f(r)}, r^2 K^{lm}(r), r^2(\sin\theta)^2 K^{lm}(r)\right)Y^{lm}(\theta,\varphi),\label{eq:deltagPolar}\ee
\be\delta g_{\mu\nu}^{(axial)}=-\frac{h_0^{lm}(r)}{\sin\theta}\frac{\partial Y^{lm}(\theta,\varphi)}{\partial \varphi}\left(\delta_{\mu}^0\delta_\nu^2+\delta_\mu^2\delta_\nu^0\right)+h_0^{lm}(r)\sin\theta\frac{\partial Y^{lm}(\theta,\varphi)}{\partial \theta}\left(\delta_{\mu}^0\delta_\nu^3+\delta_\mu^3\delta_\nu^0\right),\label{eq:deltagAxial}\ee
\be\delta A_\mu^{(polar)}=\left(\frac{u_1^{lm}(r)}{r} , 0,0,0\right)Y^{lm}(\theta,\varphi),\ee
\be \delta A_{\mu}^{(axial)}=\left(0,0,\frac{u_4^{lm}(r)}{\sin\theta}\frac{\partial Y^{lm}(\theta,\varphi)}{\partial\theta}, \frac{u_4^{lm}(r)}{\sin\theta}\frac{\partial Y^{lm}(\theta,\varphi)}{\partial\varphi}\right),\ee
\be \delta \Phi^{(polar)}=\frac{\delta \Phi^{lm}(r)}{r}Y^{lm}(\theta,\varphi),\label{eq:expansion_phi}\ee
where $Y^{lm}(\theta,\varphi)$ are the spherical harmonics, and the sum over $l$, the multipole moment, and $m$, the azimuthal number, is implicit. Given the spherical symmetry of the background, $m$ is degenerate, allowing it to be set to zero without any loss of generality. Additionally, modes with different values of $l$ are independent and thus decouple.
With the perturbation expressions in place, solving the linearized field equations (reported in App. \ref{app:equations}) allows us to fully determine the linearly perturbed spacetime metric and fields. The tidal deformation of the RBH can then be extracted from the asymptotic behavior of the spacetime metric, by comparing it with the following multipole moment expansions \cite{PhysRevD.Love}:
\be g_{00}=-1+\frac{2 m(r)}{r}+\left(\frac{2}{r^{l+1}}\left[\sqrt{\frac{4\pi}{2l+1}}\mathcal{M}_{lm} Y^{lm}(\theta,\varphi)+\left(l'<l \;\text{pole}\right)\right]-\frac{2}{l(l-1)}r^l [\mathcal{E}_{lm}Y^{lm}(\theta,\varphi)+\left(l'<l\; \text{pole}\right)] \right), \label{eq:expansion_g00}\ee
\be g_{03}=\frac{2}{r^l}\left[\sqrt{\frac{4\pi}{2l+1}}\frac{\mathcal{S}_{lm}}{l}\sin\theta \frac{\partial Y^{lm}(\theta,\varphi)}{\partial \theta}+\left(l'<l \;\text{pole}\right)\right]+\frac{2 r^{l+1}}{3 l (l-1)}\left[\mathcal{B}_{lm} \sin\theta \frac{\partial Y^{lm}(\theta,\varphi)}{\partial \theta}+\left(l'<l \;\text{pole}\right)\right],\label{eq:expansion_g03}\ee
where the sum over the harmonic indices $l,m$ is implicit and $l'<l $ pole indicates the contribution of multipoles of order $l'$ with $l'<l$. Note that in the geometries under consideration, the inclusion of matter fields introduces additional logarithmic terms and terms with irrational powers of $r$ into these expansions. These terms do not alter our definition of Love numbers and are therefore excluded in the following discussion. Note however, that the logarithmic terms, when multiplied by the same power of $r$ that defines the Love number, can be interpreted in the effective field theory framework as running of the Love number \cite{Hui:2020xxx,Kol:2011vg, Ivanov:2022hlo}.  Physically this means that the tidal deformability measured by an observer will depend on the distance at which the measure is performed (in this contest the distance is a proxy of the energy scale in the renormalization group running). Consequently the complete, running love number will not be given by a constant but will depend logarithmically on the radial distance $k_{lm} \rightarrow k0_{lm} \log(r/r_0)$ with $r_0$ a renormalization scale to be fixed by experiments.
$\mathcal{E}_{lm}$ and $\mathcal{B}_{lm}$ represent the amplitudes of the polar and axial components of the external tidal field, respectively, while $\mathcal{M}_{lm}$ and $\mathcal{S}_{lm}$ denote the mass and current multipole moments of the RBH. Mass moments (even) depend only on the polar component of the tidal field, whereas current moments (odd) depend solely on the axial component. In linear perturbation theory, these deformations are directly proportional to the applied tidal field. The tidal Love numbers $k_{lm}$ can be derived as:
\be k_{lm}^{(polar)}=-\frac{l(l-1)}{M^{2l+1}}\sqrt{\frac{4\pi}{2l+1}} \frac{\mathcal{M}_{lm}}{\mathcal{E}_{lm}},\ee \be k_{lm}^{(axial)}=3\frac{(l-1)}{M^{2l+1}}\sqrt{\frac{4\pi}{2l+1}} \frac{\mathcal{S}_{lm}}{\mathcal{B}_{lm}}.\ee
We will focus on the dominant quadrupolar mode, therefore we fix $l=2$ and, as already stated, we can take $m=0$ without loss of generality. So, we aim to compute:
\be k_{20}^{(polar)}=-\frac{2}{M^5}\sqrt{\frac{4\pi}{5}}\frac{\mathcal{M}_{20}}{\mathcal{E}_{20}}, \qquad k_{20}^{(axial)}=\frac{3}{M^5}\sqrt{\frac{4\pi}{5}}\frac{\mathcal{S}_{20}}{\mathcal{B}_{20}}.\ee
By comparing \cref{eq:expansion_g00,eq:expansion_g03} with the asymptotic expansion of $\delta g_{00}^{(polar)}$ and $\delta g_{03}^{(axial)}$ in \cref{eq:deltagPolar,eq:deltagAxial}, we can determine the values of $\mathcal{M}_{20}/\mathcal{E}_{20}$ and $\mathcal{S}_{20}/\mathcal{B}_{20}$ through the asymptotic relations:
\be-f(r)H_0^{20}(r)=-r^2 [\mathcal{E}_{20}+...]+\frac{2}{r^{3}}\left[\sqrt{\frac{4\pi}{5}}\mathcal{M}_{20}+...\right],\ee
\be h_0^{20}(r)=\frac{r^3}{3}\left[\mathcal{B}_{20}+...\right]+\frac{2}{r^2}\left[\sqrt{\frac{\pi}{5}}\mathcal{S}_{20}+...\right],\ee
where the dots indicate the additional terms in the expansions. Assuming that the additional terms on the right-hand side of each equation do not contribute any $1/r^3$ or $1/r^2$ terms (for the first and second equations, respectively), we can determine the Love numbers by analyzing the asymptotic expansion of $H_0^{20}(r)$ and $h_0^{20}(r)$. Setting $M=1$, the Love numbers are defined as follows: for polar gravitational perturbations, the Love number is given by the ratio between the coefficient of the $1/r^3$ term and that of the $r^2$ term in the asymptotic expansion of $-f(r)H_0^{20}(r)$. Similarly, for axial gravitational perturbations, it is given by the ratio between the coefficient of $1/r^2$ and that of the $r^3$ term in the asymptotic expansion of $h_0^{20}(r)$. However, this definition is not entirely straightforward, as it requires distinguishing which terms in the expansions at infinity of $-f(r)H_0^{20}(r)$ and $h_0^{20}(r)$ are due to the external tidal field and which arise from the induced response. In other worlds, verifying that the additional terms do not contribute to the $1/r^3$ or $1/r^2$ terms is not trivial.
In the next section, we discuss these ambiguities and  the calibration we adopt.

\subsection{Ambiguities and calibration}
\label{sec:ambiguities}
Describing relativistic tidal responses involves subtleties due to the ambiguous division of a perturbed metric into an external tidal part and an induced response part \cite{Fang:2005qq,Gralla:2017djj,Poisson:2020vap,Katagiri:2024wbg}. Despite this ambiguity, the overall tidally deformed metric remains well-defined once an inner boundary condition is specified. {Indeed, the metric is the solution of a second order differential equation and thus it depends on two integration constants.} Imposing an inner boundary condition {(in a BH spacetime this corresponds to imposing regularity of the metric at the horizon)} fixes one of the two integration constants in the general solution, uniquely determining the functional form of the metric components. Consequently, the description of any dynamics on the tidally deformed background is clear and unambiguous. {However, this boundary condition is not sufficient to understand which terms in the
expansion at infinity of the overall metric correspond to the induced response and
which ones correspond to the external tidal field. This division is ambiguous and rely on a specific calibration \cite{Gralla:2017djj}. However,} once a common convention for this separation is established, all results become directly comparable, even in the absence of a matching scheme between the computed tidal response function and observables such as gravitational waves. Various unified calibration schemes have been proposed in the literature. \\
In our calibration, when solving the equations analytically as a perturbative expansion around $\ell=0$ within the test-field approach, we impose regularity and continuity of the complete solution at each order in $\ell$, from the horizon to infinity. Additionally, we normalize the solution so that the dominant term at large distances is independent of $\ell$, ensuring that the Schwarzschild result is recovered at leading order in the limit $r \rightarrow \infty$. {Then, in the expansion at infinity of the overall solution, we associate every power of r smaller then $-l$ to the induced response and the remaining terms to the external tidal field:
\begin{align}
        \lim_{r\rightarrow\infty} \psi(r)&= \sum_{\alpha} r^{\alpha} = \psi_{tidal}(r)+\psi_{deform}(r) \quad \text{with}\\ 
    \psi_{tidal}= r^{l+1}&\left(1+O\left(\frac{1}{r}\right)\right) \quad \psi_{deform}= \frac{1}{r^l} \left(1+O\left(\frac{1}{r}\right)\right)
    \label{division}
\end{align}
}
This calibration aligns with that used in \cite{Cardoso:2024log}.\\
For the full gravitational problem, we impose regularity and continuity conditions on both the gravitational and matter perturbations from the horizon to infinity. The dominant term at infinity for gravitational perturbations is normalized to 1, while the parameter associated with tidal electromagnetic and scalar fields is set to zero, as we focus solely on the response of RBHs to a purely tidal \textit{gravitational} field. 
{For the division of the overall solution in the expansion at infinity we use a calibration analogous to the one used for test fields in Eq.~\eqref{division}, indeed we associate to the induced response every power of r smaller than $-l$ for axial perturbations and smaller than $-(l+1)$ for polar perturbations, as explained in Sec.~\ref{sec:tidal-deformability}.}

\section{Numerical method}
\label{sec:numerical-method}
Since a direct analytical solution of the coupled governing equations (Einstein, nonlinear Maxwell, and Klein-Gordon equations) is not feasible, we employ a numerical approach. Analytical results are only obtainable in the study of tidal deformability for test fields, as reported in Sec.~\ref{sec:test}.
To obtain numerical solutions of the equations of motion, we use direct integration techniques ~\cite{Chandrasekhar:1975zza,Pani:2013pma}. In this section, we explain the method in detail. We first determine the behavior of gravitational, electromagnetic, and scalar perturbations at the horizon and at infinity, expressing them in terms of expansion parameters. These parameters are then fixed by ensuring continuity of the perturbation functions and their derivatives, allowing us to extract the asymptotic expansion of the gravitational perturbations and compute the Love number. \\
At the horizon, located at $r_h$, we impose regularity on a given perturbation $\gamma(r)$ by performing a Taylor expansion:
 \be
\gamma(r)\xrightarrow{r\to r_h}\sum_{i=0}^\infty \gamma_i \,(r-r_h)^i,
\ee
where $\gamma_i$ are constants. Substituting this into the governing equations and solving order by order up to a certain $i$, we obtain an expansion near $r_h$ for the perturbations. To ensure the convergence of the result at the end of the numerical integration (see App. \ref{app:num-analysis}), we truncate the sum at $i=5$. The number of free parameters is two for purely electromagnetic sources and three when a scalar field is involved.\\
At infinity, we first solve the equations analytically for $M=\ell=0$, finding that each perturbation is a combination of two power-law solutions. In general, these exponents can be irrational for matter perturbations, and due to the coupling of the equations, such exponents may also appear in the gravitational perturbation solution for $M\not=0$ and $\ell\geq 0$. However, this does not hinder the numerical solution, as we systematically account for them. For $M\not=0$ and $\ell\geq 0$, the asymptotic expansion we consider is \cite{Cardoso:2024log}:
\be
\gamma(r)\xrightarrow{r\to\infty}C(r)\left(\sum_{k=0}^\infty \frac{\Tilde{\gamma}_k^{(0)} + \Tilde{\gamma}_k^{(1)}\log\left(\frac{r}{2M}\right)}{r^k}\right)+D(r)\left(\sum_{k=0}^\infty \frac{\hat{\gamma}_k^{(0)} + \hat{\gamma}_k^{(1)} \log\left(\frac{r}{2M}\right)}{r^k}\right),
\label{eq:asymptexp}
\ee
where $\Tilde{\gamma}_k^{(0)},\,\Tilde{\gamma}_k^{(1)},\,\hat{\gamma}_k^{(0)}$ and $\hat{\gamma}_k^{(1)}$ are constants. The function $C(r)$ corresponds to the dominant gravitational perturbation, while $D(r)$  corresponds to the subdominant matter field perturbation, both obtained in the $M=\ell=0$ case (see Sec. \ref{sec:ambiguities} for details on this choice). If the matter perturbation contains only integer powers of $r$, the second term (involving $D(r)$) is effectively absorbed into the first, reducing \cref{eq:asymptexp} to a single-term expansion. Logarithmic terms arise due to matter fields. If they contribute to the Love number, their coefficients play a role analogous to a $\beta$-function in renormalization flow, suggesting a \textit{running} Love number \cite{Hui:2020xxx,Kol:2011vg, Ivanov:2022hlo}. In the Schwarzschild case, the analytical solution for tidal perturbations also includes a logarithmic term, but it is excluded by imposing regularity at the horizon. In contrast, in our case, such terms may persist due to the presence of matter, even after imposing regularity conditions. We therefore include them in the asymptotic expansion at infinity to account for the full solution.
As we will show, these terms generally do not vanish (see App. \ref{app:asymptotic-numerical-sol}). We solve the equations order by order up to $k=5$, which guarantees the convergence and stability of the results, {as shown in App.~\ref{app:num-analysis}}.
We normalize the gravitational tidal perturbation parameter to 1, and with the choice described in \cref{eq:asymptexp}, we automatically set to zero the parameters associated with external electromagnetic and scalar tidal fields. With these choices, the number of free parameters at infinity is two for purely electromagnetic sources and three when a scalar field is present. \\
Finally, we determine the parameters using a shooting method. Specifically, we use the parametric expansions at infinity and at the horizon as initial conditions. We then integrate the equations numerically, solving from infinity back to a matching point $r_\text{cut}$ and from just outside the horizon to $r_\text{cut}$. This provides the perturbations values at $r_\text{cut}$ in terms of one set of parameters when solving from the horizon, and in terms of another set of parameters when solving from infinity. We then enforce the continuity of the perturbations and their first derivatives at $r_\text{cut}$, allowing us to determine the parameters values. After obtaining the curves, we perform a fit to extrapolate their behavior. The errors in the fitting coefficients are estimated to be of order 1 in the final decimal place, and the procedure for obtaining them is described in App.~\ref{app:num-analysis}.

\section{Test-field perturbations}\label{sec:test}
Before presenting the results for gravitational perturbations of RBHs, we first analyze test-field perturbations on top of these spacetimes. While less physically relevant, this approach allows us to study tidal responses without making assumptions about the matter distribution sustaining the geometry.
We begin by studying the equations of motion for the test fields perturbatively in $\ell$ (with $0\le\ell\ll1$), using generic metric functions $m(r)$ and $\phi(r)$. This generic parametrized approach helps identify the conditions on the metric functions necessary for a nonzero Love number. Next, we focus on the specific choices of $m(r)$ and $\phi(r)$ that correspond to the RBH models under consideration, deriving analytical solutions for the test fields perturbatively in $\ell$. These specific results are then compared with the generic analysis to ensure consistency. Additionally, we numerically solve the equations--following the approach outlined in Sec.~\ref{sec:numerical-method}--for the full range of $\ell$ that defines an object with an event horizon. For small $\ell$, the numerical results are cross-checked with the analytical ones to validate our numerical method. \\
The equation governing static scalar, electromagnetic and gravitational test fields in the spacetime described by \cref{metric} reads~\cite{Karlovini:2001fc,Medved:2003pr}:
\be
\frac{\dd^2 \psi_s}{\dd r_*^2}  - V_s\psi_s = 0\,,
\label{HarmEq2}
\ee
where $\psi_s$ represents the spin-$s$ perturbation, and the tortoise coordinate $r_*$ is defined as $\dd r_*/\dd r \equiv \e^{\phi}/f$. The potential $V_s$, that depends on the spin-weight of the perturbation and on the metric functions, is given by the following expression:
\be
V_s = f\,\e^{-2\phi} \left[\frac{l(l+1)}{r^2}+\frac{2 \left(1-s^2\right) m}{r^3}-(1-s)
\left(\frac{2 m'}{r^2}+\frac{f \phi'}{r}\right)\right].
\label{potential}
\ee
We begin our analysis of tidal deformability for test fields in the following section, where we use generic metric functions $m(r)$ and $\phi(r)$.

\subsection{Parametrized approach at leading order}
\label{sec:GeneralAnalysis}
In \cite{Cardoso:2024log}, a parametrized formalism was developed to compute the Love numbers of static and spherically symmetric black hole backgrounds. This approach assumes that deviations from the Schwarzschild case in the perturbation equations can be absorbed into additional terms in the potential, without modifying the definition of the tortoise coordinate relative to the Schwarzschild case. This assumption is justified because, as shown in \cite{Cardoso:2019mqo}, through a suitable field redefinition, equations for scalar and vector fields on a static, spherically symmetric spacetime close to Schwarzschild can always be cast in this form.
However, in most RBH models, the required field redefinition becomes singular at $r=2M$, leading to singular behavior in the additional terms of the potential. As a result, regularity cannot be imposed at the horizon for the redefined field but only for the original, physical field. To address this issue, we present a new parametrized analysis of the test-field tidal response of spherically symmetric black holes, expressed in terms of the expansions of $m(r)$ and $\phi(r)$. The two parametrized approaches yield identical results in cases where the aforementioned field redefinition remains non-singular, as in the Simpson-Visser RBH model. {This is because, for the metric models where the field redefinition introduced in \cite{Cardoso:2024log} is regular, imposing regularity of the redefined field is equivalent to imposing regularity of the original physical field. In these cases, in order to recover the results in terms of our new parametrization from the results in terms of the parametrization introduced in \cite{Cardoso:2024log}, is sufficient to express the coefficients of the series expansion of the potential in Eq.~\eqref{potential} used in \cite{Cardoso:2024log} in terms of the coefficients of the series expansions of $m$ and $\phi$ that we are going to use.} We now explicitly present results for $s=l=2$, before generalizing them at the end of the section. 

\subsubsection{Case $\phi=0$}
First, we analyze the case where the only non-vanishing metric function is $m(r)$, corresponding to a Bardeen-like spacetime, as discussed in Sec.~\ref{sec:RBHs}.
We solve \cref{HarmEq2} perturbatively in $\ell$, imposing regularity at the horizon at each order.
We expand the mass function as a power series in $\ell$\footnote{This polynomial expansion holds for most proposed models but is not universally valid; see, for instance, the Dymnikova model \cite{dymnikova_vacuum_1992}.}:
\begin{equation}
    m(r)=M+\sum_{i=1}^{\infty}  \ell^i \sum_{k=1}^{Ni}M_{ik} \frac{ M^{k-i+1}}{r^k},
    \label{massExp}
\end{equation}
where $M_{ik}$ are dimensionless parameters. 
Let $j$ denote the leading order in $\ell$ at which corrections appear in the mass function  ($m(r)=M+\ell^j \sum_{k=1}^{Nj} M_{jk}\frac{M^{k-j+1}}{r^k}+ O(\ell^{j+1})$). At each order, the general solution will contain two free parameters. We constrain these by imposing regularity at $r=2M$ and setting to 0 the coefficient of $r^3$ for the solution at orders $j>0$ in $\ell$. These coefficients correspond to a renormalization of the zeroth-order horizon-regular solution. 
To obtain a nonzero $1/r^2$ term (and thus a nonzero Love number) at order $\ell^j$, $Nj$ must be at least 4. The resulting solution is:
\begin{align}
\psi(r) \propto & 
 \text{ } r^3\left[ 1+\ell^j \left(\frac{4}{3r^2} M_{j1} M^{2-j} +\frac{8  M^{3-j}(2  M_{j1}+3  M_{j2})}{9r^3} +\frac{4 M^{4-j}(2  M_{j1}+3  M_{j2}+18  M_{j3})}{3 r^4}\right)\right] +\nonumber\\
& +\frac{M^{5-j} \ell^j}{r^2} \left[\frac{32 M_{j4} (1+5 \log \left(\frac{r}{2M}\right))}{25}+\sum_{k=5}^{Nj} \frac{2^{7-k}  k  M_{jk} }{5(-4+k) }\right]  \left( 1+\mathcal{O}\left(\frac{1}{r}\right)\right) + \mathcal{O}\left(\ell^{j+1}\right).
\end{align} 

\subsubsection{Case $\phi \neq0$}
In the general case, we introduce an additional metric function, $\phi(r)$, which we expand perturbatively in $\ell$:
\begin{equation}
    \phi(r)=\sum_{i=1}^{\infty}  \ell^i \sum_{k=1}^{Ri} \phi_{ik}\frac{ M^{k-i}}{r^k},
\end{equation}
where $\phi_{ik}$ are dimensionless parameters. The mass function is also expanded as in \cref{massExp}.
Let $j$ be the leading order in $\ell$ for corrections to both the mass function in \cref{massExp} and $\phi$. If these two functions have different leading orders, say $p$ and $q$, we define $j$ as the smallest of the two. To ensure a nonzero coefficient for the $1/r^2$ term (i.e., a nonzero Love number) at order $\ell^j$, at least one of $Rj$ or $Nj$ must be at least 4. The resulting perturbation solution is:
\begin{align}
\psi(r)& \propto r^3\left[ 1+\ell^j \left(\frac{\phi_{j1}M^{1-j}}{r}+\frac{M^{2-j}(4M_{j1}-\phi_{j1}+\phi_{j2})}{3r^2}  +\frac{2 M^{3-j} (8  M_{j1}+12 M_{j2}-2  \phi_{j1}-4 \phi_{j2}+9 \phi_{j3})}{9r^3} \right. \right.+\nonumber\\
& \left. \left. +\frac{M^{4-j}(8   M_{j1}+12  M_{j2}+18  M_{j3}-2 \phi_{j1}-4  \phi_{j2}-\phi_{j3}+12 \phi_{j4})}{3 r^4}\right)\right]
+\frac{M^{5-j} \ell^j }{r^2}\left[\frac{32 (M_{j4}+ \phi_{j4})\left(1+5 \log \left(\frac{r}{2M}\right)\right)}{25}+\right.
\nonumber\\
& \left.+\frac{ 4\phi_{j5}\left(-4+5 \log \left(\frac{r}{2M}\right)\right)}{5}+\sum_{k=5}^{Nj} \frac{2^{7-k}  k  M_{jk} }{5(-4+k) }+\sum_{k=6}^{Nj} \frac{2^{7-k}  k  \phi_{jk} }{5(20 - 9 k + k^2) }\right]  \bigg( 1+\mathcal{O}\left(\frac{1}{r}\right)\bigg) + \order{\ell^{j+1}}.
\label{testfieldGeneral}
\end{align}  
Even though specific numerical coefficients may vary, this structure remains valid for general values of $l$ ans $s$. Specifically, at leading order $j$ in $\ell$, a coefficient $M_{jn}$ or $\phi_{j p}$ in the metric functions leads to the following scenarios:
\begin{itemize}
    \item \textit{Zero Love numbers} (no terms proportional to $1/r^l$) if $n< 2s$ and $p< 2s$ ($n< 2$ and $p< 2$ if $s=0$).
    \item \textit{Nonzero running Love numbers} (terms proportional to $1/r^l$ and $\log[r]/r^l$) if $2s \leq n\leq 2l$ or $2s \leq p\leq 2l+1$ ($2 \leq n\leq 2l$ or $2 \leq p\leq 2l+1$ if $s=0$).
    \item \textit{Nonzero Love numbers} (terms proportional to $1/r^l$) if $ n> 2l$ or $ p> 2l+1$.
\end{itemize}
At higher orders in $\ell$, mixing terms between different $M_{jk}$ and $\phi_{jk}$ appear, making the general solution increasingly complex. However, we explicitly computed these solutions for specific cases in the following section.
  
\subsection{Beyond the leading order for specific models}\label{fit}
Now, we compute the test-field tidal deformations for specific RBH cases with $s=l=2$, extending our analysis beyond the leading order in $\ell$. We also compare the analytical results with numerical computations to validate our numerical method. {As explained in App.~\ref{app:num-analysis}, we report numerical coefficients such that the numerical uncertainty is of the same order as the last digit shown.}
\begin{figure}[ht]
    \begin{subfigure}{0.4\textwidth}
        \centering
        \textbf{Bardeen}\par\medskip
        \includegraphics[width=\linewidth]{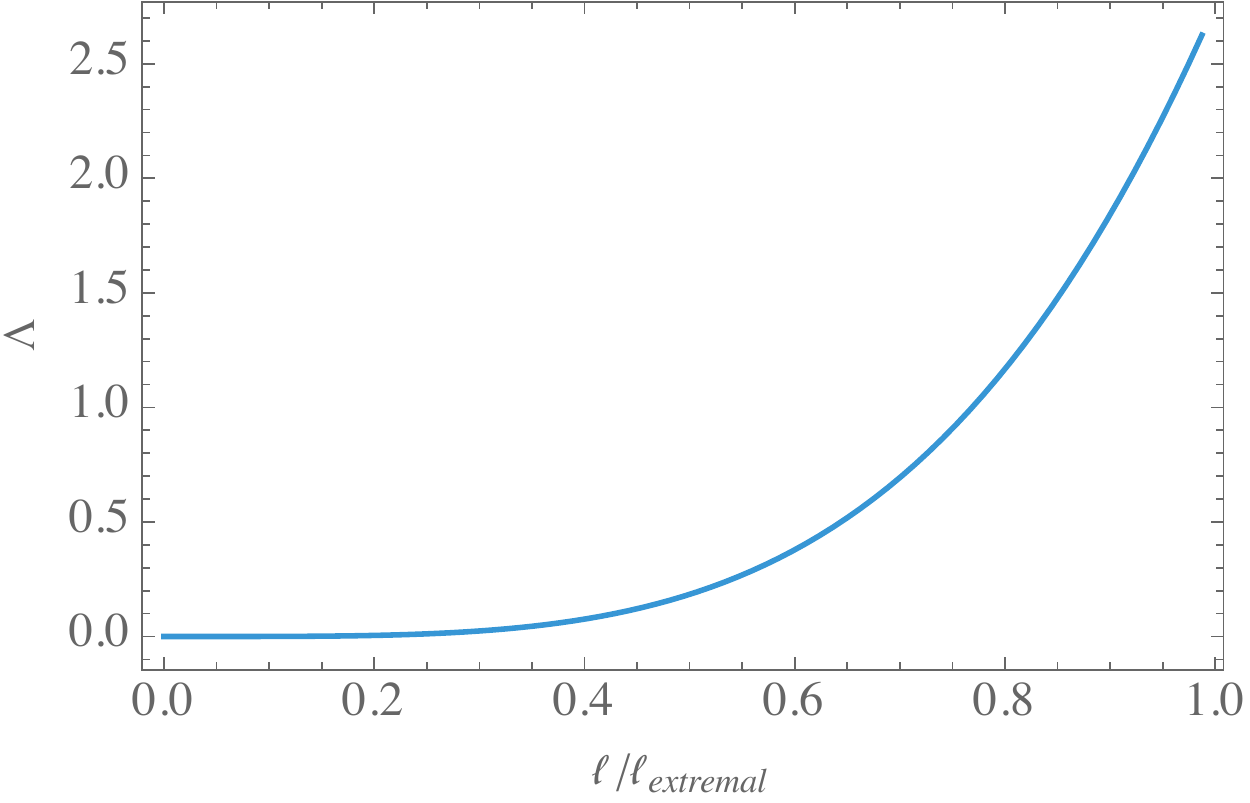}
    \end{subfigure}
   \hspace{2mm}
    \begin{subfigure}{0.4\textwidth}
        \centering
        \textbf{Hayward}\par\medskip
        \includegraphics[width=\linewidth]{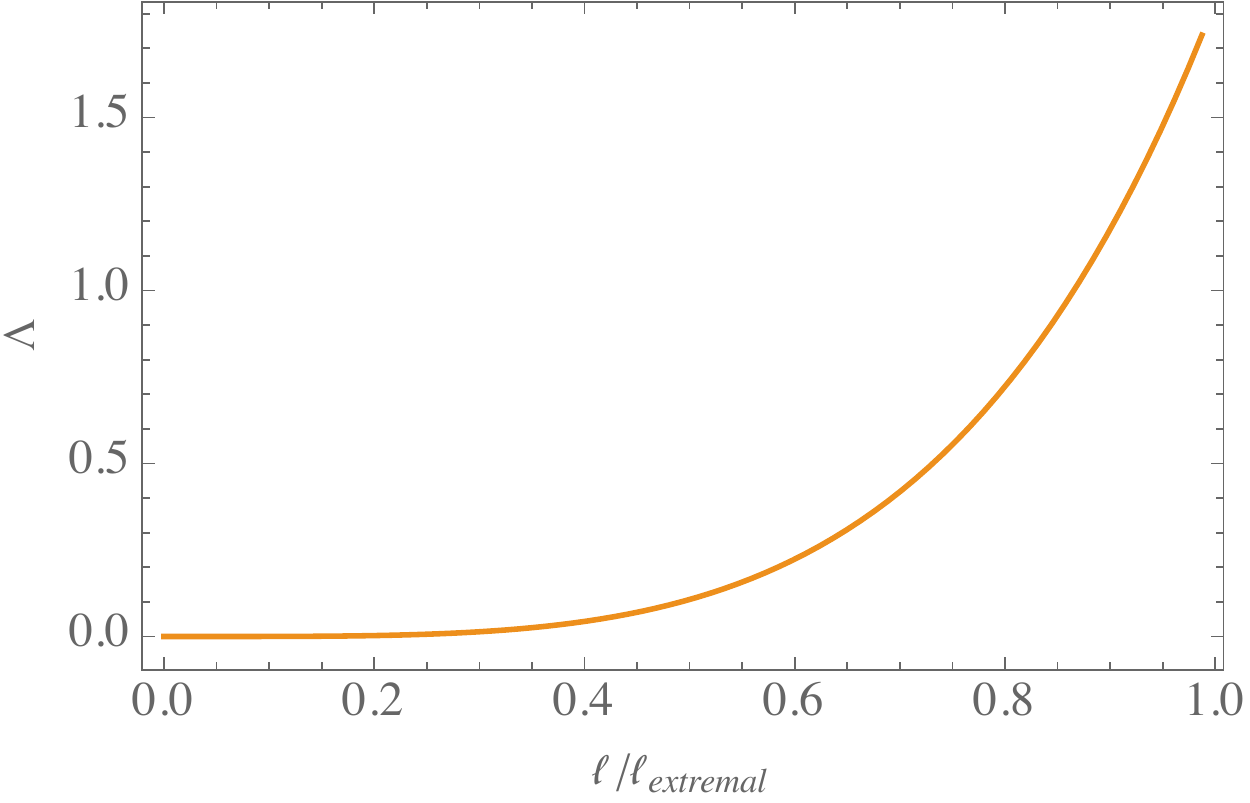}
    \end{subfigure}
\par\bigskip
    \begin{subfigure}{0.42\textwidth}
        \centering
        \textbf{Fan-Wang}\par\medskip
        \includegraphics[width=\linewidth]{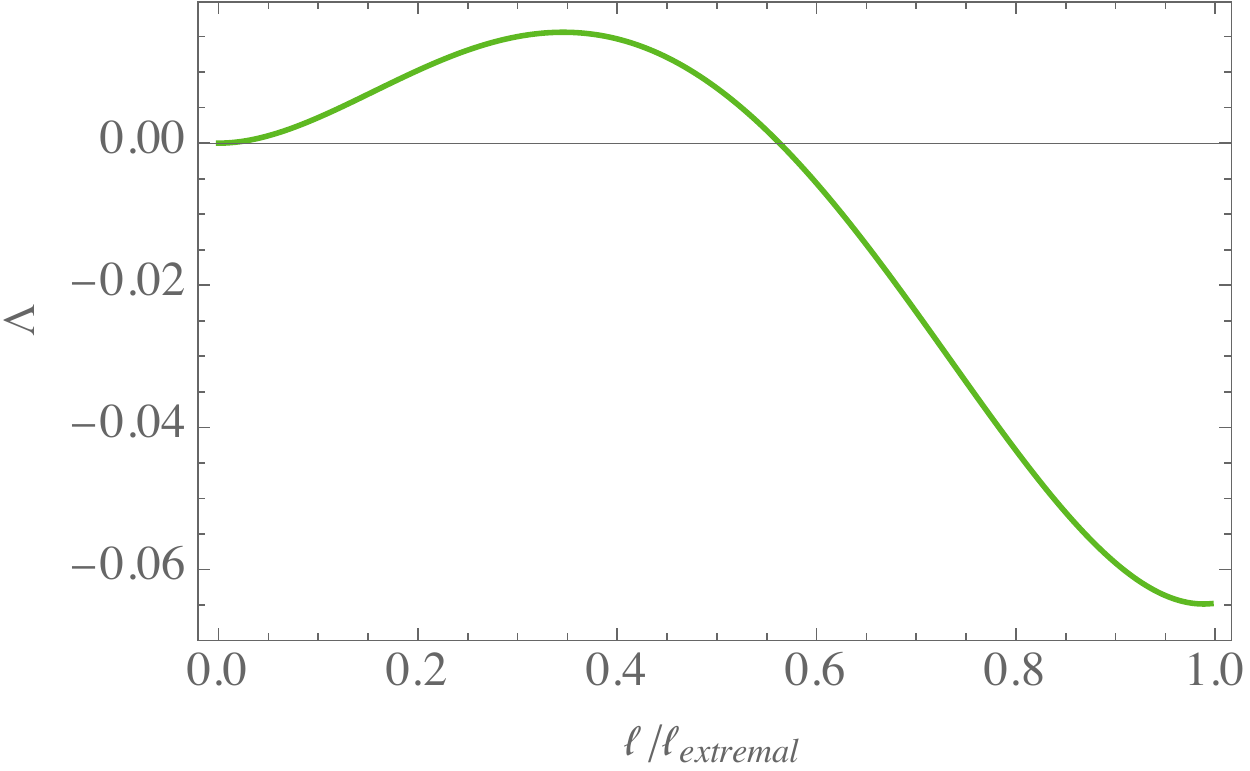}
    \end{subfigure}
  \hspace{2mm}
    \begin{subfigure}{0.4\textwidth}
        \centering
        \textbf{Simpson-Visser}\par\medskip
        \includegraphics[width=\linewidth]{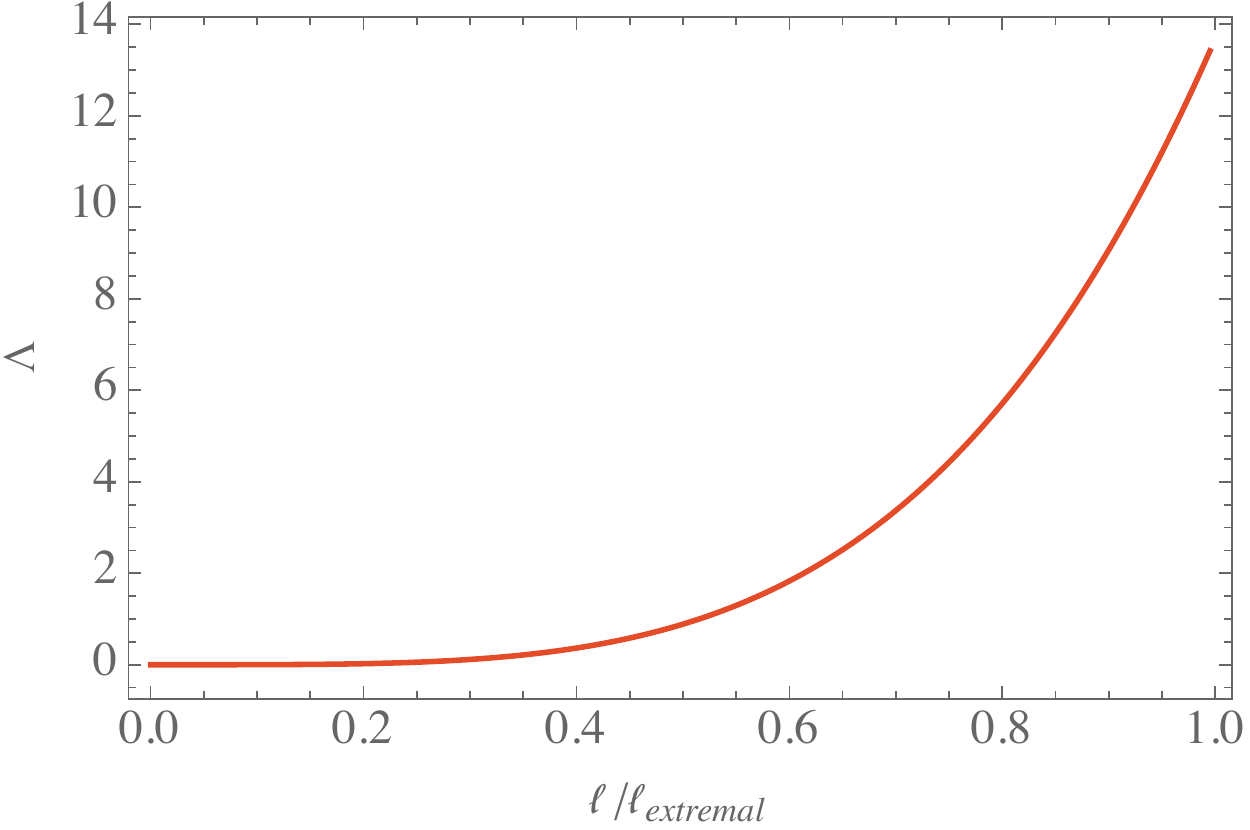}
    \end{subfigure}
 \caption{\textit{Test-field perturbations.} The plots show the value of $\Lambda$, defined as the ratio of the coefficient of the $1/r^2$ term to that of the $r^3$ term in the asymptotic expansion of $\psi(r)$, as a function of the normalized regularization parameter $\ell/\ell_{extremal}$ for each RBH model. Here, $\psi(r)$ represents the $s=l=2$ test-field perturbation introduced on top of each RBH metric. All plots are shown for $M=1$.}
    \label{fig:test-field}
\end{figure}\\
By solving \cref{HarmEq2} analytically up to $\mathcal{O}(\ell^4)$ for Bardeen, Hayward and Simpson-Visser, and up to $\mathcal{O}(\ell^2)$ for Fan-Wang, and expanding the result at infinity, we obtain:
\be\textit{Bardeen:}\qquad \label{eq:analytical-Bardeen-testfield}
\psi(r) &\propto   r^3-4 M \ell^2-\frac{6 M^2 \ell^2}{r}+\frac{6 M \ell^4\left(7+10 \log \left(\frac{r}{2M}\right)\right)}{5 r^2}+\order{\frac{1}{r^3}},\ee
\be\textit{Hayward:}\qquad \psi(r)\propto r^3-\frac{12M^2 \ell^2}{r}+\frac{24 M \ell^4}{5 r^2}+\order{\frac{1}{r^3}}
\label{Hay},\ee
\be\textit{Fan-Wang:}\qquad  \psi(r) &\propto r^3-4 M \ell r-\frac{16}{3}M \ell(M-3 \ell) -\frac{4 M^2 \ell(2 M-15 \ell)}{r}+\frac{128 M^3 \ell^2\left(1+5 \log \left(\frac{r}{2M}\right)\right)}{25 r^2}
+\mathcal{O}\left(\frac{1}{r^3}\right)
\label{eq:analytical-FW-testfield},\ee
\be\textit{Simpson-Visser:}\qquad \psi(r) &\propto  r^3-\frac{4 M \ell^2}{3}-\frac{2 M^2 \ell^2}{r}+\frac{4 M \ell^4\left(17-15\log \left(\frac{r}{2M}\right)\right)}{75 r^2}+\order{\frac{1}{r^3}}.\label{eq:analytical-SV-testfield}\ee
Solving the same equation numerically and fitting the coefficient of $1/r^2$ as a function of $\ell$ (setting the coefficient of $r^3$ to 1), we obtain (see Fig. \ref{fig:test-field}):
\be\textit{Bardeen:}\;\Lambda\sim 8.4 M\ell^4 
\qquad\textit{Hayward:}\; \Lambda\sim4.8 M \ell^4
\qquad\textit{Fan-Wang:}\; \Lambda\sim5.1 M^3\ell^2 
\qquad\textit{Simpson-Visser:}\; \Lambda=0.9 M\ell^4 
\ee
The fitted coefficients from numerical results are in perfect agreement with the analytical predictions.\\
To compare the analytical results with the previous analysis of general $m(r)$, we expand $m(r)$ at leading order in $\ell$ (see Eqs.~\eqref{eq:m(r)-Bardeen}-\eqref{eq:m(r)-SV}): 
\be\textit{Bardeen:}\qquad m_B(r)=M-\frac{ 3 \ell^2 M}{2 r^2}+\mathcal{O}(\ell^4),\ee
\be\textit{Hayward:}\qquad m_H(r)= M-\frac{ 2 \ell^2 M^2}{r^3}+\mathcal{O}(\ell^4),\ee
\be\textit{Fan-Wang:}\qquad m_{FW}(r)=M-\frac{3\ell M}{r}+\mathcal{O}(\ell^2),\ee
\be\textit{Simpson-Visser:}\qquad m_{SV}(r)=M+\frac{(-2M+r)\ell^2}{2 r^2} \quad  \text{and} \quad 
\phi_{SV}(r)= -\frac{\ell^2}{2 r^2}+\order{\ell^4}.\ee
For the Bardeen metric, the leading-order correction to $m(r)$ is $\mathcal{O}(\ell^2)$, corresponding to a term $M_{22} M/r^2$ in \cref{massExp}. Since $n=2<2s=4$, the general analysis predicts a zero Love number at $\ell^2$, in agreement with \cref{eq:analytical-Bardeen-testfield}. Similarly, for the Hayward metric, the leading-order correction to $m(r)$ is $\mathcal{O}(\ell^2)$, corresponding to $M_{23} M^2/r^3$. This term leads to a zero Love number in \cref{Hay} and a $1/r^2$ term appears only at $\mathcal{O}(\ell^4)$, this confirms that for $n=3<2s=4$, the Love number remains zero at leading order in $\ell$. For the Fan-Wang metric, the leading-order correction to $m(r)$ is $\mathcal{O}(\ell)$, corresponding to $M_{11}M/r$. Since $n=1<2s=4$, the Love number is predicted to be zero at order $\ell$, in agreement with \cref{eq:analytical-FW-testfield}. Finally, for the Simpson-Visser metric, at leading order in $\ell$, the only nonzero coefficients in \cref{testfieldGeneral} are $M_{2,n=1}$, $M_{2,n=2}$ and $\phi_{2,p=2}$. Since $n,p<4$, we expect a zero Love number at $\ell^2$, which is confirmed in \cref{eq:analytical-SV-testfield}, as it appears only at order $\ell^4$. \\
For the reader's reference, we note that increasing $\ell$ beyond the extremal black hole value--thus exploring horizonless configurations--results in a continuous extension of the numerical Love number curve. The curve smoothly transitions from the RBH case, through the extremal case, to the horizonless case, despite the change in boundary conditions: regularity at the horizon is replaced by regularity at the origin.\\
This concludes our analysis of test-field perturbations. In the next section, we extend our study to the full gravitational system, employing the same numerical method that we used and validated with test fields.

\section{Full gravitational perturbations}
\label{sec:resultsfullgrav}
\begin{figure}[ht]
    \begin{subfigure}{0.49\textwidth}
    \centering
        \textbf{Polar}\par\medskip
        \includegraphics[width=\linewidth]{Polar-Total.pdf}
    \end{subfigure}
    \hfill
    \begin{subfigure}{0.49\textwidth}
    \centering
        \textbf{Axial}\par\medskip
        \includegraphics[width=\linewidth]{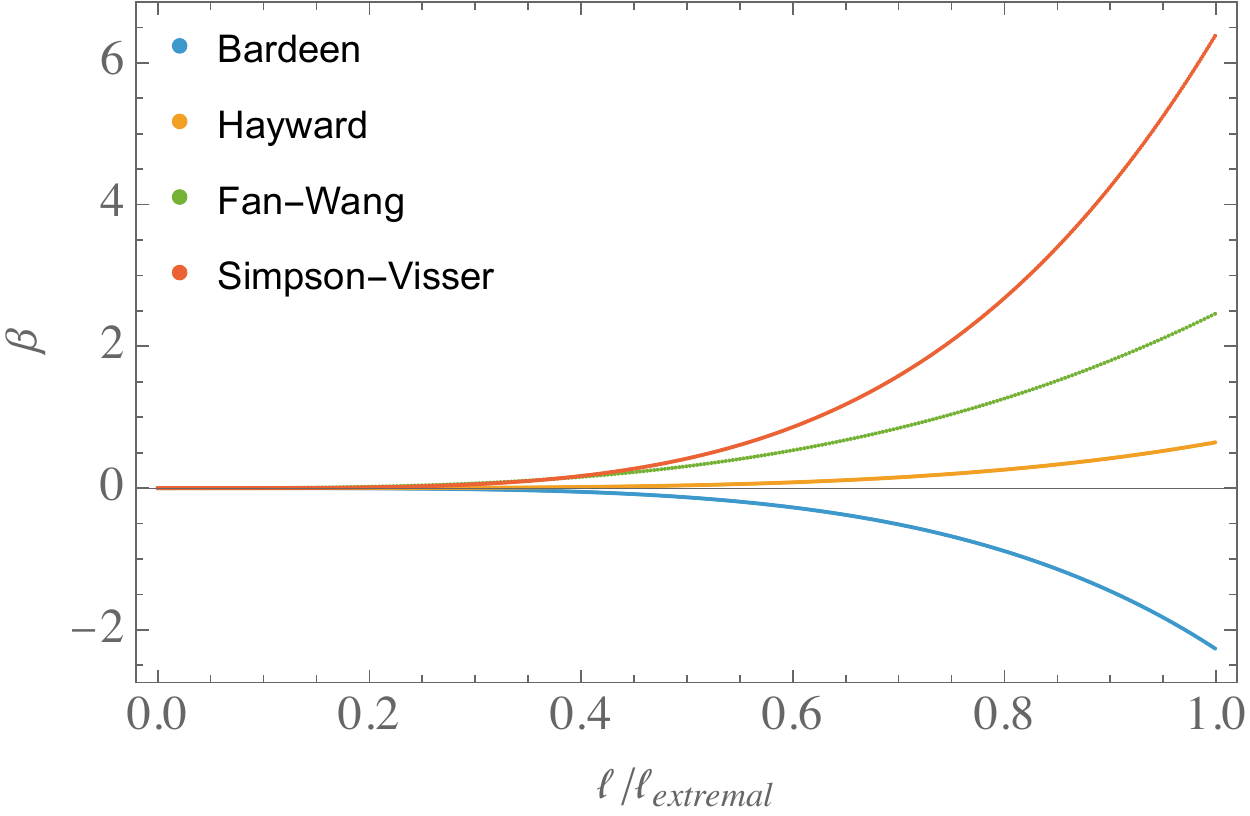}
    \end{subfigure}
     \caption{\textit{Polar (left) and axial (right) gravitational perturbations.} The plots depict the value of $\alpha$ ($\beta$), the coefficient of the $1/r^3$ ($1/r^2$) term in the asymptotic expansion of $-f(r) H_0^{20}(r)$ ($h_0^{20}(r)$), as a function of the normalized regularization parameter $\ell / \ell_{\text{extremal}}$ for each RBH model, as specified in the legends. All plots are shown for $M=1$.}
     \label{fig:Love}
\end{figure}

Now, we analyze the tidal response of RBHs in the full gravitational system. For each parity sector, gravitational, scalar and electromagnetic harmonic perturbations satisfy a system of coupled, non-homogeneous wave equations, which can be written schematically as
\be
\frac{\dd^2\mathcal{I}}{\dd r_*^2} -V_\mathcal{I} \mathcal{I}
+ \sum_{\mathcal{J\neq I}} c_\mathcal{I,J}\,\mathcal{J} &= 0\,,\label{eqpert_gen}
\ee
where $r_*$ is the tortoise coordinate, defined as $\dd r_*/\dd r \equiv \e^{\phi}/f$. The indices $\mathcal{I,J}$ take the values $\{A,E\}$ in the sector where axial gravitational perturbations couple to polar electromagnetic perturbations, and $\{P,B,S\}$ in the sector where polar gravitational, axial electromagnetic and polar scalar perturbations are coupled. The variables $\{\mathcal{A,P,B,E,S}\}$ represent specific combinations of the metric, the electromagnetic potential, and the scalar field perturbation functions and their derivatives. The potentials $V_\mathcal{I}$ and the coefficients $c_\mathcal{I,J}$ are determined by the background metric and fields and depend on the harmonic number $l$ from the spherical-harmonics expansion. The explicit forms of these equations are provided in App. \ref{app:equations}.\\
Solving these equations analytically via a perturbative expansion in $\ell$ around the Schwarzschild solution is particularly challenging. 
In the axial sector, we can solve the equations up to $\mathcal{O}(\ell^2)$, where we find that the Love numbers for all the RBH models remain zero, except for the Bardeen model. For the latter, it is not possible to extend the analytical solution beyond order $\ell^0$, preventing us from determining its tidal Love number (recall that order 0 in $\ell$ corresponds to Schwarzschild field equations in the gravitational sector). In the polar sector, the situation is even more restrictive: for all the RBH models considered, analytical solutions cannot be obtained beyond order $\ell^0$. Nevertheless, for both parity sectors, we employ the numerical approach described in Sec. \ref{sec:numerical-method}, previously applied in the test-field approximation, to numerically compute the tidal deformations of RBHs up to the extremal RBH value of $\ell$, which represents the largest value for which the model retains an horizon. Numerical results for the asymptotic expansion coefficients of the gravitational perturbation--directly related to the Love number, as discussed in Sec.~\ref{sec:tidal-deformability}--are presented below for both polar and axial gravitational perturbations. In \cref{fig:Love}, we show the coefficients of the $1/r^3$ and $1/r^2$ terms (for polar and axial perturbations, respectively) in the asymptotic expansion of the gravitational perturbation induced by an external tidal field. The analytical asymptotic behavior of the gravitational perturbations up to $1/r^5$ is provided in App. \ref{app:asymptotic-numerical-sol}.\\

\subsubsection{Polar gravitational perturbations} 
As discussed in Sec.\,\ref{sec:tidal-deformability}, to study the tidal deformability of RBHs under polar gravitational perturbations, we focus on the coefficient of the $1/r^3$ term in the asymptotic expansion of $-f(r)H_0^{20}(r)$. The numerical results for this coefficient are shown in the left panel of Fig.~\ref{fig:Love} for each RBH model, with $\ell$ varying from 0 to the extremal RBH value. As expected, due to the presence of matter fields, the asymptotic behavior includes not only a term of the form $\alpha/r^3$, where $\alpha$ is the coefficient of interest, but also a logarithmic contribution $\Tilde{\alpha}\log (r/2M)/r^3$ for all models. The coefficient $\Tilde{\alpha}$ can be interpreted as a beta function in the context of a classical renormalization flow, signifying the running of the Love number \cite{Hui:2020xxx,Kol:2011vg, Ivanov:2022hlo}. The plot of $\alpha$ for each RBH model is shown in Fig.~\ref{fig:Love}, while the coefficient $\Tilde{\alpha}$ is given in the asymptotic expansions of App.~\ref{app:asymptotic-numerical-sol}.
In the limit $\ell\to0$, we find $\alpha\to0$ and $\Tilde{\alpha}\to 0$, as expected, since in this limit we recover the Schwarzschild solution, which has a vanishing Love number \cite{Binnington:2009bb}.\\
For the Bardeen model, $\alpha$ follows the trend $\sim \text{a}\, M^3\ell^2  + \text{b} \,M\ell^4 
$, with a numerical fit yielding $\text{a} \simeq 0.03$ and $ \text{b}\simeq 10$. The Simpson-Visser model exhibits a similar behavior, with $\alpha\sim \text{a}\, M^3\ell^2 + \text{b} \,M\ell^4 
$, where $\text{a}\simeq 0.05$ and $\text{b}=40$. For the Hayward model, we find $\alpha\sim \text{a} M^3\ell^2 
$, with $\text{a}\,\simeq 13$. 
Finally, for the Fan-Wang model, $\alpha\sim \text{a}\, M^2\ell^3 
$, with $a\simeq 2.2\times 10^2$. For details on the fitting procedure, see App.~\ref{app:num-analysis}.

\subsubsection{Axial gravitational perturbations}
The tidal deformability of RBHs under axial gravitational perturbations is analyzed by examining the coefficient $\beta$ of the $1/r^2$ term in the asymptotic expansion of $h_0^{20}(r)$, as outlined in Sec.~\ref{sec:tidal-deformability}. The right panel of Fig~\ref{fig:Love} presents the numerical results for this coefficient across the various RBH models, with $\ell$ ranging from 0 to the extremal RBH limit. As in the polar case, the coefficients of the logarithmic terms are provided in the asymptotic expansions of App.~\ref{app:asymptotic-numerical-sol}. Notably, the Hayward model lacks a logarithmic term, meaning its axial Love number does not exhibit a running behavior. 
The coefficient $\beta$ follows different trends for each RBH model (see App.~\ref{app:num-analysis} for details on the fitting procedure): Bardeen: $\beta\sim \text{a} \,M\ell^4 
$, with $\text{a} \simeq -5.8$; Hayward: $\beta\sim \text{a}\, M\ell^4 
$, with $\text{a} \simeq 1.8$; Fan-Wang:  $\beta\sim \text{a} \,M^2\ell^3 
$, with $\text{a} \simeq 95$; Simpson-Visser: $\beta\sim \text{a}\,M \ell^4 
$, with $\text{a} \simeq 0.4$. In both the Hayward and Simpson-Visser models, the positivity and increase of the polar and axial gravitational Love numbers with $\ell$ are features that also emerge in the test-field approach. \\
A summary of the dominant behavior of the Love numbers for test-field, axial, and polar gravitational perturbations is provided in Table \ref{summary}. The trends described there remain a good approximation for the entire signal up to $\ell_{\text{extremal}}$.
\begin{table}[]
\centering
\begin{tabular}{@{}cccc@{}}
\toprule
               & Test-field & Axial & Polar \\ \midrule
Bardeen        &     $42/5 M \ell^4 $ & $ \sim -5.8 M \ell^4$     &    $ \sim 0.03 M^3 \ell^2 +10 M \ell^4$       \\
Hayward        &     $24/5 M \ell^4$       &   $\sim 1.8 M \ell^4 $    & $\sim 13 M^3 \ell^2 $      \\
Fan-Wang       &      $128/25 M^3 \ell^2$      &  $\sim 95 M^2 \ell^3 $    &   $\sim 220 M^2 \ell^3 $         \\
Simpson-Visser &     $68/75 M \ell^4 $       &     $\sim 0.4 M \ell^4 $   &  $\sim 0.05 M^3 \ell^2 +40 M \ell^4$    \\ \bottomrule
\end{tabular}
\caption{\textit{Love numbers.} The table presents the dominant coefficients in the perturbative expansion of the $l=2$ Love numbers at small $\ell$. The first column contains exact analytical coefficients for test-field perturbations, while the axial and polar results in the last two columns are obtained numerically.}
\label{summary}
\end{table}

\subsection{Detection}
We now consider a system of two RBHs with masses $M_1$ and $M_2$ merging, and we ask whether this event can be distinguished from a merger of two singular black holes through the phase shift induced by their nonzero Love numbers. We define the total mass of the system as $m=M_1+M_2$ and the reduced mass as $\mu=\frac{M_1 M_2}{M1+M_2}$. 
As shown in \cite{LNinGW}, within the stationary phase approximation, a tidal Love number modifies the phase $\Psi(f)$ of the Fourier transform of the gravitational wave signal at the frequency $f=\omega/\pi$ by:
\be \delta \Psi(f)=- \frac{9}{16} \frac{v^5}{\mu M^4}\left[\left(11 \frac{M_1}{M_2}+\frac{m}{M_1}\right)\lambda_1 +1\leftrightarrow 2\right],\label{eq:deltaFreq}\ee
where $v=(\pi m f)^{1/3}$ and $\lambda_1$ ($\lambda_2$) is the tidal deformability of the object with mass $M_1$ ($M_2$). For simplicity, we assume $M_1=M_2=m/2$. 
The highest frequency at which we can evaluate the phase shift due to tidal deformability in Eq.~\eqref{eq:deltaFreq} is the innermost stable circular orbit frequency $f_c$ of the object. Using the expression for merging black holes as an approximation, this frequency is given by \cite{Iscofreq}: 
\be f_c \sim \frac{2}{\pi 6^{3/2} m}.\ee
Thus, the phase shift at $f_c$ becomes:
\be \delta\Psi (f_c)=- \frac{9}{4} \frac{ (\pi m f_c)^{5/3}}{m^5}\left[2\left(11+2\right)\lambda \right]\sim -2.1 \frac{\lambda}{m^5},\ee
where $\lambda$ is the Love number. The values of $\delta\Psi (f_c)$ for each model are summarized in Table \ref{table2}.
\begin{table}[]
\centering
\begin{tabular}{@{}ccc@{}}
\toprule
&   $\delta\Psi_{Axial}$ [rad] & $\delta\Psi_{Polar}$ [rad] \\ \midrule
Bardeen   &  $ \sim 6.1\,(\ell/m)^4$     &    $ \sim -0.08\, (\ell/m)^2 -10\, (\ell/m)^4$       \\
Hayward        &     $\sim -1.9\, (\ell/m)^4 $    & $\sim -3.4\, (\ell/m)^2 $      \\
Fan-Wang        &  $\sim -50\, (\ell/m)^3 $    &   $\sim -115\, (\ell/m)^3 $         \\
Simpson-Visser  &     $\sim -0.4\, (\ell/m)^4 $   &  $\sim -0.01\, (\ell/m)^2 -42\, (\ell/m)^4$    \\ \bottomrule
\end{tabular}
\caption{\textit{Phase shifts}. The table shows the phase shifts contributions in the Fourier transform of the gravitational wave signal from an equal-mass RBH merger due to the dominant terms of the $l=2$ Love numbers, for both axial (left) and polar (right) gravitational perturbations. Here, $m$ denotes the total mass of the binary RBH merger.}
\label{table2}
\end{table}
The phase shift depends on the ratio $\ell/m$, with different scaling behaviors for each model and perturbation's polarity. If $\ell$ is of the order of the Planck length--much smaller than typical astrophysical masses--the shift is negligible. However, if $\ell$ approaches its extremal value (of order $m$), the phase shift reaches approximately one radian.  While this is significantly lower than phase shifts observed in binaries neutron star mergers ($\delta \Psi\sim 10^3\, \text{rad}$), it remains potentially measurable by third-generation gravitational wave detectors \cite{Punturo:2010zz, LISA:2017pwj}.

\section{Conclusions}
\label{sec:concl}
In this work, we explored the tidal response of regular black holes (RBHs) to external test-field and gravitational perturbations. Our study focused on four models, representative of the two possible families of RBHs. \\
For both test-field and gravitational perturbations, we found a nontrivial tidal response, in contrast to the zero-response of the classical Schwarzschild black hole (see Fig. \ref{fig:test-field} and \ref{fig:Love}). The tidal Love numbers depend on the mass $M$ and the regularizing parameter $\ell$, an additional parameter characterizing these regular spacetimes. The precise scaling of the tidal response with $\ell$ varies depending on the specific model and the type of perturbation. 
A summary of these dependencies is provided in Table \ref{summary}.\\
For the test-field tidal response, we developed and applied a parametrized formalism in which the Love numbers of a generic spherically symmetric regular (or deformed) black hole are expressed in terms of the perturbative expansion in $\ell$ of the two metric functions. This allows for an analytical, albeit perturbative, computation of the Love numbers. \\
For what regards gravitational tidal response instead, we solve the field equations numerically using a direct integration shooting method. 
As expected, the tidal Love numbers increase in magnitude with $\ell$ and vanish in the Schwarzschild limit ($\ell=0$). In all cases, they grow as a power law in $\ell$. {Even though the largest values of the Love numbers are obtained for extremal configurations ($\ell = \ell_{\text{extremal}}$), we do not attribute this behavior to the extremal nature of the black hole horizon itself. Rather, we believe it arises from the fact that these values of $\ell$ correspond to the biggest deviations from the Schwarzschild metric. Nevertheless, it is interesting to note that there is some evidence suggesting that quantum or beyond-GR corrections to classical black holes--and to their tidal deformability in particular--are more significant for extremal rotating black holes \cite{Horowitz:2023xyl}.
}\\
{As summarized in Table \ref{summary} and Fig.~\ref{fig:Love}, both the test-field and full gravitational perturbations exhibit the same qualitative trend: all Love numbers grow in magnitude with a power law of $\ell$ and reach $\mathcal{O}(1-10)$ in the extremal limit. However, the exact power laws and numerical coefficients generically differ because our matter source couples to both axial and polar gravitational sectors. 
Nonetheless, we observe that the Hayward and Simpson-Visser models have very similar behavior across the test-field and full axial gravitational perturbations: they share the same leading dependence on $M$ and $\ell$, and their deformation coefficients are of comparable magnitude. This suggests that, in these models, the specific matter content--or the theoretical interpretation of the model--has less impact on the tidal response of the RBH. Note also that, although the Fan-Wang coefficients reported in Table \ref{summary} may appear large at first glance, the extremal value of $\ell$ in this model is smaller than in the others (see \cref{eq:m(r)-Bardeen,eq:m(r)-FW,eq:m(r)-Hayward,eq:m(r)-SV}), and as a result, the maximal tidal deformation remains comparable across all configurations (see Fig.~\ref{fig:Love}). We also observe that the axial-sector Love number in the Bardeen model (and the Fan-Wang one in the test-field approach) is negative. It would be worthwhile to explore the implications of this behavior in future work, as it may offer insights into the physical viability of these models. Finally, we note that Fan-Wang model exhibits nearly identical behavior in both the axial and polar gravitational perturbations. Additionally, the polar perturbations of the Bardeen and Simpson-Visser models display similar scaling and coefficients; however, the extremal value $\ell_\text{extremal}$ for the Simpson-Visser model is more than twice that of the Bardeen case, leading to a larger maximal tidal deformability.}
We remark that while our analysis was conducted within general relativity coupled to {specific} exotic matter, {given the qualitative agreement of results for test-fields and gravitational perturbations,} we expect that a nonzero Love number could also emerge in a hypothetical modified-quantum gravity theory that gives rise to RBH solutions. \\
Although a direct connection between our computed Love numbers and gravitational wave observables remains ambiguous, we estimate that the resulting phase shift in binary black hole mergers could reach a few radians. 
This suggests that RBH tidal effects may be detectable by third-generation interferometers. In this sense, gravitational wave observations could potentially offer a glimpse into the interior structure of black holes, helping to address open questions and missing pieces in our understanding of gravity.\\
A key direction for future work is to compute the Love numbers of these objects using an effective action approach to verify consistency with our method and resolve ambiguities. Once established, the next step would be a detailed study of how RBHs modify the gravitational wave signal from binary mergers, providing potential observational signatures of the interior of a black hole. 

\section*{Acknowledgments}
We thank Edgardo Franzin for the useful insights, the interesting discussions and for his relevant comments on an early draft of this manuscript.
C.C. and V.V. thank Perimeter Institute for hospitality where parts of this work were carried out.
This work was supported by the Natural Sciences and
Engineering Research Council of Canada (NSERC).
L.L. also thanks financial support via
the Carlo Fidani Rainer Weiss Chair at Perimeter Institute and CIFAR. This research was supported in part by Perimeter Institute for Theoretical Physics. Research at Perimeter Institute is supported in part by the Government of Canada through the Department of Innovation, Science and Economic Development and by the Province of Ontario through the Ministry of Colleges and Universities. C.C. thanks the ISSNAF organization for its financial support.

\appendix
\section{Explicit perturbative equations}\label{app:equations}
Using the perturbations of the metric $g_{\mu\nu}$, the electromagnetic potential $A_\mu$, and the scalar field $\Phi$ as defined in Eqs. ~\eqref{eq:deltagPolar}-\eqref{eq:expansion_phi}, we derive the equations of motion for linear perturbations from Eqs.~\eqref{eq:modMaxwell}, \eqref{eq:Einstein}, and \eqref{eq:KG}.  \\ 
\textit{Axial gravitational perturbations:}

\begin{subequations}\be
&f u_1'' + f \left(\phi'-\frac{2 \ell^2 \mathcal{L}_{FF}}{r^5 \mathcal{L}_F}\right) u_1' + \left(\frac{2 \ell^2 f \mathcal{L}_{FF}}{r^6 \mathcal{L}_F}-\frac{r f \phi'+l(l+1)}{r^2}\right) u_1 - \frac{l (l+1) \ell}{r^3}\,h_0 =0\,,\label{sec1max1.static}\\
&f h_0'' + f \phi' h_0' -\frac{2 r^2 f \left(r \phi'+1\right)+(l-1)(l+2) r^2+4 \ell^2 \mathcal{L}_F}{r^4}\,h_0  - \frac{4 \ell \mathcal{L}_F}{r^3}\,u_1 = 0\,.\label{AxialEq13.static}
\ee
\end{subequations}
\textit{Polar gravitational perturbations:}
\be
&\delta\Phi''
-\left(\phi'-\frac{f'}{f}\right)\delta\Phi'
+\left(\frac{\phi'}{r} - \frac{l(l+1) + r f' - r^2 V_{\Phi\Phi}}{r^2 f}\right)\delta\Phi
-\frac{r V_\Phi H_0}{f}
+\frac{r(2H_0'+2K')}{2}\,\Phi' =0\,,\label{KGsect2_a.static}\\
&u_4''
+\left(\frac{f'}{f}-\frac{2 \ell^2 \mathcal{L}_{FF}}{r^5
\mathcal{L}_F}-\phi'\right) u_4'
- \frac{l (l+1)}{r^2 f}\left(1  + \frac{\ell^2 \mathcal{L}_{FF}}{r^4 \mathcal{L}_F}\right)u_4
+\frac{\ell}{r^2 f}\left(1+\frac{\ell^2 \mathcal{L}_{FF}}{r^4 \mathcal{L}_F}\right)K = 0
\,.\label{Maxwellsector2.static}
\ee
\be
f H_0'' - \eta_1 H_0' - \eta_2 H_0 + J_3 = 0\,,
\ee
where 
the coefficients are defined as:
\be
\eta_1 &= \frac{4\ell^2}{\Delta}\left(f'-2 f\phi'\right)\left(\mathcal{L}_F+\frac{\ell^2\mathcal{L}_{FF}}{r^4}\right)-f'+f\left(\phi'-\frac{2}{r}\right),\\
\eta_2 &= \frac{4\ell^2}{\Delta}\left(\mathcal{L}_F+\frac{\ell^2\mathcal{L}_{FF}}{r^4}\right)\left(\frac{l (l+1)}{r^2}-4 f'\phi'+\frac{f'^2}{f}+\frac{2 f\left(r\phi'-1\right)\left(2 r\phi'+1\right)}{r^2}\right) - 2 f'\left(2\phi'+\frac{1}{r}\right)\nonumber\\
&\phantom{=}+\frac{f'^2}{f}+\frac{2 f\left(r\phi'+1\right)\left(2 r\phi'-1\right)}{r^2}+\frac{l^2+l+2}{r^2}-\frac{4\ell^2\mathcal{L}_F}{r^4}\,,\\
J_3 &= -\frac{4\ell}{r^2}\left[\frac{2 f\mathcal{L}_F}{r}-\left(f'-\frac{2 f\left(r\phi'-1\right)}{r}\right)\left(\frac{4\ell^4\mathcal{L}_F\mathcal{L}_{FF}}{\Delta r^4}+\frac{4\ell^2\mathcal{L}_F^2}{\Delta}+\mathcal{L}_F\right)\right] u_4'
-\frac{4 \ell (l-1)l (l+1)(l+2)}{r^2 \Delta}\left(\mathcal{L}_F +\frac{\ell^2\mathcal{L}_{FF}}{r^4}\right)u_4\nonumber\\
&\phantom{=} -\frac{32\ell^2 f\sqrt{\pi r\phi'}}{r^2\Delta}\left(\mathcal{L}_F+\frac{\ell^2\mathcal{L}_{FF}}{r^4}\right)\delta\Phi'\nonumber\\
&-\frac{4}{r}\left[\frac{8\ell^2}{\Delta}\left(\mathcal{L}_F+\frac{\ell^2\mathcal{L}_{FF}}{r^4}\right)\left(\sqrt{\pi r\phi'}\left(\frac{f'}{r}+\frac{f\left(1-2 r\phi'\right)}{r^2}\right)-2\pi V_\Phi\right)-\frac{\sqrt{\pi} f\left[r\phi''+\phi'\left(2 r\phi'+3\right)\right]}{r\sqrt{r\phi'}}
\right]\delta\Phi\,,\\
\Delta &= (l-1)(l+2)r^2 + 4\ell^2\mathcal{L}_F\,,
\ee
and the metric function $K$ is given by:
\be
K = -\frac{r^2 f\left[l(l+1) - 4 r^2 f'\phi'+2 f\left(r\phi'-1\right)\left(2 r\phi'+1\right)\right] + r^4 f'^2}{\Delta f}\,H_0
-\frac{r^4 \left(f'-2 f\phi'\right)}{\Delta}\,H_0'
+\frac{4\ell l (l+1) \mathcal{L}_F}{\Delta}\,u_4\nonumber\\
+\frac{4 \ell r \mathcal{L}_F \left[r f'+2f(1-r\phi')\right]}{\Delta}\,u_4'
+\frac{16\pi r^3 V_\Phi + 8 r\sqrt{\pi r\phi'}\left[f\left(2 r\phi'-1\right)-r f'\right]}{\Delta}\,\delta\Phi
-\frac{8 r^2 f\sqrt{\pi r\phi'}}{\Delta}\,\delta\Phi'\,.
\ee

\section{Asymptotic numerical solutions}
\label{app:asymptotic-numerical-sol}
We present the asymptotic behavior of the gravitational perturbations obtained numerically for each RBH model, considering both polar and axial gravitational perturbations. The coefficient of the external gravitational tidal field term is normalized to 1, while the coefficients corresponding to tidal matter fields are set to 0. The parameters $\beta$ and $\delta$ are free parameters derived from the asymptotic expansion, and are determined using a shooting method in conjunction with the solution at the outer horizon.\\
\textbf{Bardeen:}\\
\textit{Polar gravitational perturbations:}
    \begin{align*}
H_0^{20}(r)\sim &\,r^2-2 M r + \frac{3 M \ell^2}{r}  - \frac{M^2 \ell^2}{2 r^2} 
+ \frac{r^{1-\sqrt{10}} \big(-30 M \ell \beta + 15 \sqrt{10} M \ell \beta \big) + \delta + \frac{8}{5} M^3 \ell^2 \log \left(\frac{r}{2M}\right) + 2 M \ell^4 \log \left(\frac{r}{2M}\right)}{r^3}+ \\
&+ \frac{\frac{31 M^4 \ell^2}{15} + \frac{253 M^2 \ell^4}{12} + r^{1-\sqrt{10}} \big(60 M^2 \ell \beta - 6 \sqrt{10} M^2 \ell \beta\big) + 3 M \delta + \frac{24}{5} M^4 \ell^2 \log \left(\frac{r}{2M}\right) + 6 M^2 \ell^4 \log \left(\frac{r}{2M}\right)}{r^4}+ \\
&+ \frac{\frac{926 M^5 \ell^2}{147} + \frac{6488 M^3 \ell^4}{147} - \frac{995 M \ell^6}{24} + r^{1-\sqrt{10}} \big(\frac{600}{13} M^3 \ell \beta + \frac{216}{13} \sqrt{10} M^3 \ell \beta\big) + \frac{50 M^2 \delta}{7} + \frac{80}{7} M^5 \ell^2 \log \left(\frac{r}{2M}\right) + \frac{100}{7} M^3 \ell^4 \log \left(\frac{r}{2M}\right)}{r^5}.
\end{align*}
\textit{Axial gravitational perturbations:}
\begin{align*}
h_0^{20}(r)\sim &\,r^3 -2 M r^2 + 3 M \ell^2 + r^{-1-\sqrt{6}} \left( -\frac{6}{5} M \ell \beta + \frac{9}{5} \sqrt{6} M \ell \beta \right) 
+ \frac{r^{-\sqrt{6}} \left( \frac{22284 M^2 \ell \beta}{2185} - \frac{1926 \sqrt{6} M^2 \ell \beta}{2185} \right) + \delta + \frac{72}{25} M \ell^4 \log \left(\frac{r}{2M}\right)}{r^2} +\\
&+ \frac{\frac{177 M^2 \ell^4}{25} + r^{-\sqrt{6}} \left( \frac{103644 M^3 \ell \beta}{437 (12 + 7 \sqrt{6})} + \frac{40284 \sqrt{6} M^3 \ell \beta}{437 (12 + 7 \sqrt{6})} - \frac{42 M \ell^3 \beta}{12 + 7 \sqrt{6}} - \frac{21 \sqrt{\frac{3}{2}} M \ell^3 \beta}{12 + 7 \sqrt{6}} \right) + \frac{4 M \delta}{3} + \frac{96}{25} M^2 \ell^4 \log \left(\frac{r}{2M}\right)}{r^3} +\\
&+ \frac{\frac{6966 M^3 \ell^4}{1225} - \frac{257 M \ell^6}{40} + r^{-\sqrt{6}} \left( \frac{164088 M^4 \ell \beta}{5609 + 2294 \sqrt{6}} + \frac{66988 \sqrt{6} M^4 \ell \beta}{5609 + 2294 \sqrt{6}} - \frac{211209 M^2 \ell^3 \beta}{5 (5609 + 2294 \sqrt{6})} - \frac{86687 \sqrt{6} M^2 \ell^3 \beta}{5 (5609 + 2294 \sqrt{6})} \right) + \frac{40 M^2 \delta}{21} + \frac{192}{35} M^3 \ell^4 \log \left(\frac{r}{2M}\right)}{r^4} \\
&+ \frac{\frac{4737 M^4 \ell^4}{1225} - \frac{19227 M^2 \ell^6}{1000} + \frac{20 M^3 \delta}{7} + \frac{3}{2} M \ell^2 \delta + \frac{288}{35} M^4 \ell^4 \log \left(\frac{r}{2M}\right) + \frac{108}{25} M^2 \ell^6 \log \left(\frac{r}{2M}\right)}{r^5}.
\end{align*}
\textbf{Hayward:}\\
\textit{Polar gravitational perturbations:}
\begin{align*}
H_0^{20}(r)\sim &r^2 -2 M r + \frac{34 M^2 \ell^2}{7 r^2} + \frac{\delta - \frac{1248}{245} M^3 \ell^2 \log \left(\frac{r}{2M}\right)}{r^3} + \frac{-\frac{2692}{245} M^4 \ell^2 + r^{\frac{3}{2} - \frac{\sqrt{57}}{2}} \left( -\frac{864}{7} M^2 \ell \beta + \frac{144}{7} \sqrt{57} M^2 \ell \beta \right) + 3 M \delta - \frac{3744}{245} M^4 \ell^2 \log \left(\frac{r}{2M}\right)}{r^4}+ \\
&+ \frac{-\frac{73352 M^5 \ell^2}{2401} - \frac{1262 M^3 \ell^4}{49} + r^{\frac{3}{2} - \frac{\sqrt{57}}{2}} \left( \frac{36909}{98} M^3 \ell \beta - \frac{3015}{98} \sqrt{57} M^3 \ell \beta \right) + \frac{50 M^2 \delta}{7} - \frac{12480}{343} M^5 \ell^2 \log \left(\frac{r}{2M}\right)}{r^5}.
\end{align*}
\textit{Axial gravitational perturbations:}
\begin{align*}
h_0^{20}(r)\sim  &\,r^3-2 M r^2  + \frac{4 M^2 \ell^2}{r} + \frac{\delta}{r^2}+ \frac{4 M \delta}{3 r^3}  + \frac{-\frac{4}{735} M^2 \left( 1767 M \ell^4 - 350 \delta - 945 \ell \beta \right) - \frac{648}{35} M^3 \ell^4 \log \left(\frac{r}{2M}\right)}{r^4}+ \\
&+ \frac{-\frac{2}{245} M^3 \left( 2985 M \ell^4 - 350 \delta - 1680 \ell \beta \right) - \frac{1728}{35} M^4 \ell^4 \log \left(\frac{r}{2M}\right)}{r^5}.
\end{align*}
\textbf{Fan-Wang}\\
\textit{Polar gravitational perturbations:}
\begin{align*}
H_0^{20}(r)\sim &\, r^2 -2 M r + 6 M \ell - \frac{12 M \ell^2}{r} + \frac{95 M \ell^3}{4 r^2} + \frac{\delta + \frac{3}{10} M \ell^3 \left( -53 M + 45 \ell \right) \log \left(\frac{r}{2M}\right)}{r^3}+ \\
&+ \frac{\frac{1}{8} M \ell^3 \left( -254 M^2 + 771 M \ell + 1239 \ell^2 \right) + 3 M \left( \delta + 8 \beta \right) + \frac{9}{10} M \ell^3 \left( -66 M^2 + 155 M \ell - 105 \ell^2 \right) \log \left(\frac{r}{2M}\right)}{r^4} +\\
&+ \left(\frac{1}{1960} M \left( -210484 M^3 \ell^3 + 611391 M^2 \ell^4 - 35 \left( 15661 \ell^6 + 48 \ell \left( 3 \delta + 100 \beta \right) \right) + 10 M \left( 61662 \ell^5 + 56 \left( 25 \delta + 312 \beta \right) \right) \right)+\right. \\
&\left.+ \frac{3}{70} M \ell^3 \left( -3664 M^3 + 12759 M^2 \ell - 17250 M \ell^2 + 7875 \ell^3 \right) \log \left(\frac{r}{2M}\right)\right)\frac{1}{r^5}.
\end{align*}
\textit{Axial gravitational perturbations:}
\begin{align*}
h_0^{20}(r)\sim  &\,r^3-2 M r^2  + 6 M r \ell - 13 M \ell^2 + \frac{2 M \ell^2 \left( M + 10 \ell \right)}{r} + \frac{\delta + \frac{3}{10} M \ell^3 \left( -218 M + 95 \ell \right) \log \left(\frac{r}{2M}\right)}{r^2} +\\
&+ \frac{\frac{1}{45} M \left( 15 \ell^3 \left( -412 M^2 - 1228 M \ell + 587 \ell^2 \right) + 60 \left( \delta + 3 \beta \right) \right) + \frac{1}{2} M \ell^3 \left( -436 M^2 - 328 M \ell + 77 \ell^2 \right) \log \left(\frac{r}{2M}\right)}{r^3} +\\
&+ \left( -\frac{M \left( 1269372 M^3 \ell^3 + 6170682 M^2 \ell^4 + 105 \ell \left( 7047 \ell^5 - 72 \delta + 180 \beta \right) - 10 M \left( 259746 \ell^5 + 168 \left( 5 \delta + 24 \beta \right) \right) \right)}{4410} +\right. \\
&\left. - \frac{1}{140} M \ell^3 \left( 59296 M^3 + 53116 M^2 \ell - 49460 M \ell^2 + 5775 \ell^3 \right) \log \left(\frac{r}{2M}\right) \right) \frac{1}{r^4} +\\
&+ \left( -\frac{1}{47040} M \left( 23553792 M^4 \ell^3 + 133385952 M^3 \ell^4 + 30 M \ell \left( 1061460 \ell^5 - 896 \left( 5 \delta - 32 \beta \right) \right) \right. \right. +\\
&\left. \left. - 35 \ell^2 \left( 237867 \ell^5 - 8832 \delta + 6720 \beta \right) - 4 M^2 \left( 32704923 \ell^5 + 33600 \left( \delta + 6 \beta \right) \right) \right) \right.+ \\
&\left. + \frac{1}{280} M \ell^3 \left( -209280 M^4 - 104496 M^3 \ell + 402116 M^2 \ell^2 - 172420 M \ell^3 + 13475 \ell^4 \right) \log \left(\frac{r}{2M}\right) \right)\frac{1}{r^5} .
\end{align*}
\textbf{Simpson-Visser:}\\
\textit{Polar gravitational perturbations:}
\begin{align*}
H_0^{20}(r)\sim&\, r^2-2 M r  - \frac{2 \ell^2}{3} - \frac{M^2 \ell^2}{10 r^2} + \frac{4 M \ell^2}{3 r} + \frac{r^{1-\sqrt{10}} \left( -6 M \ell \beta + 3 \sqrt{10} M \ell \beta \right) + \delta + \frac{8}{25} M^3 \ell^2 \log \left(\frac{r}{2M}\right) - \frac{2}{15} M \ell^4 \log \left(\frac{r}{2M}\right)}{r^3} +\\
&+ \frac{\frac{31 M^4 \ell^2}{75} - \frac{47 M^2 \ell^4}{90} + r^{1-\sqrt{10}} \left( \frac{1536}{65} M^2 \ell \beta - \frac{186}{13} \sqrt{\frac{2}{5}} M^2 \ell \beta \right) + 3 M \delta + \frac{24}{25} M^4 \ell^2 \log \left(\frac{r}{2M}\right) - \frac{2}{5} M^2 \ell^4 \log \left(\frac{r}{2M}\right)}{r^4} +\\
&+ \left( \frac{926 M^5 \ell^2}{735} + \frac{2441 M^3 \ell^4}{3150} + \frac{M \ell^6}{490} + r^{1-\sqrt{10}} \left( \frac{53388 M^3 \ell \beta}{65 (16 + 7 \sqrt{10})} + \frac{16464 \sqrt{\frac{2}{5}} M^3 \ell \beta}{13 (16 + 7 \sqrt{10})} \right. \right. +\\
&\left. \left. + \frac{39 M \ell^3 \beta}{16 + 7 \sqrt{10}} + \frac{57 \sqrt{\frac{5}{2}} M \ell^3 \beta}{16 + 7 \sqrt{10}} \right) + \frac{50 M^2 \delta}{7} + \frac{9 \ell^2 \delta}{14} + \frac{16}{7} M^5 \ell^2 \log \left(\frac{r}{2M}\right) - \frac{56}{75} M^3 \ell^4 \log \left(\frac{r}{2M}\right) - \frac{3}{35} M \ell^6 \log \left(\frac{r}{2M}\right) \right)\frac{1}{r^5} .
\end{align*}
\textit{Axial gravitational perturbations:}
\begin{align*}
h_0^{20}(r)\sim &\, r^3 -2 M r^2  + \frac{13 M \ell^2}{15} - \frac{r \ell^2}{2} + \frac{\frac{4 M^2 \ell^2}{15} - \frac{\ell^4}{8} + \frac{6 M r^{-\sqrt{6}} \ell \beta}{2 + 3 \sqrt{6}}}{r} + \frac{r^{-\sqrt{6}} \left( \frac{22284 M^2 \ell \beta}{10925} - \frac{1926 \sqrt{6} M^2 \ell \beta}{10925} \right) + \delta + \frac{2}{25} M \ell^4 \log \left(\frac{r}{2M}\right)}{r^2} +\\
& +\frac{- \frac{497}{225} M^2 \ell^4 - \frac{\ell^6}{16} + r^{-\sqrt{6}} \left( \frac{2988 M^3 \ell \beta}{2185} + \frac{1614 \sqrt{6} M^3 \ell \beta}{2185} + \frac{27}{50} M \ell^3 \beta \right)+ \frac{4 M \delta}{3} + \frac{8}{75} M^2 \ell^4 \log \left(\frac{r}{2M}\right)}{r^3} + \left( -\frac{72686 M^3 \ell^4}{11025} - \frac{293 M \ell^6}{9800} +\right.\\
&\left. +r^{-\sqrt{6}} \left( \frac{1251096 M^4 \ell \beta}{5(42696 + 17519 \sqrt{6})} + \frac{510744 \sqrt{6} M^4 \ell \beta}{5(42696 + 17519 \sqrt{6})} \right. \right.  \left. \left. + \frac{35149092 M^2 \ell^3 \beta}{475(42696 + 17519 \sqrt{6})} + \frac{14237463 \sqrt{6} M^2 \ell^3 \beta}{475(42696 + 17519 \sqrt{6})} \right) + \right.\\
&+\frac{40 M^2 \delta}{21} + \frac{4 \ell^2 \delta}{7}  \left. + \frac{16}{105} M^3 \ell^4 \log \left(\frac{r}{2M}\right) + \frac{8}{175} M \ell^6 \log \left(\frac{r}{2M}\right) \right) \frac{1}{r^4} + \left(-\frac{16536 M^4 \ell^4}{1225} - \frac{171433 M^2 \ell^6}{91875} - \frac{5 \ell^8}{128} + \frac{20 M^3 \delta}{7} +  \right.\\
& \left.+ \frac{299}{210} M \ell^2 \delta+\frac{8}{35} M^4 \ell^4 \log \left(\frac{r}{2M}\right) + \frac{299 M^2 \ell^6 \log \left(\frac{r}{2M}\right)}{2625}\right)\frac{1}{r^5}
\end{align*}
We note that in each case, the asymptotic expansion for $\ell=0$ and $\beta=0$ yields the well-known Schwarzschild solution (which is obtained before imposing regularity at the event horizon):\\
\textit{Polar gravitational perturbations for Schwarzschild:}
    \begin{equation*}
    H_0^{20}(r)\sim r^2-2 M r  + \frac{\delta}{r^3}+ \frac{3 M \delta}{r^4} + \frac{50 M^2 \delta}{7 r^5} 
\end{equation*}
\textit{Axial gravitational perturbations for Schwarzschild:}
\begin{equation*}
h_0^{20}(r)\sim  r^3-2 M r^2  + \frac{\delta}{r^2} + \frac{4 M \delta}{3 r^3}  + \frac{40 M^2 \delta}{21 r^4} + \frac{20 M^3 \delta}{7 r^5}
\end{equation*}
Only the parameter $\delta$ appears in the Schwarzschild case, as the parameter $\beta$ is associated with the exotic matter that sustains the RBH structure, and thus does not appear in the Schwarzschild solution.

\section{Numerical analysis}
\label{app:num-analysis}
{The exact numerical values obtained in the paper depend on several algorithmic choices--namely, the location of numerical infinity, the matching radius $r_\text{cut}$, the numerical horizon, and the truncation orders of the series expansions at both the horizon and infinity. To confirm stability, we systematically varied each of these parameters. 
First, we confirmed that varying the parameter $r_\text{cut}$--the matching point at which we impose continuity between the solutions integrated from the horizon and from infinity--does not affect the final outcome. Second, we examined the dependence on the parameter $\epsilon$, which determines the initial distance from the horizon at which the integration begins. Keeping the expansion near the horizon fixed to order $(r-rh)^5$, we found that decreasing $\epsilon$ does not alter the results, indicating stability with respect to the choice of starting point. Third, we investigated the convergence of the results for increasing order of the asymptotic expansion at infinity and at the horizon. As an illustration, we provide a plot showing the behavior of the solution in the Hayward model as the expansion order at infinity is varied. An analogous analysis was performed for all other models considered to ensure the convergence of the results presented.}
\begin{figure}[ht]
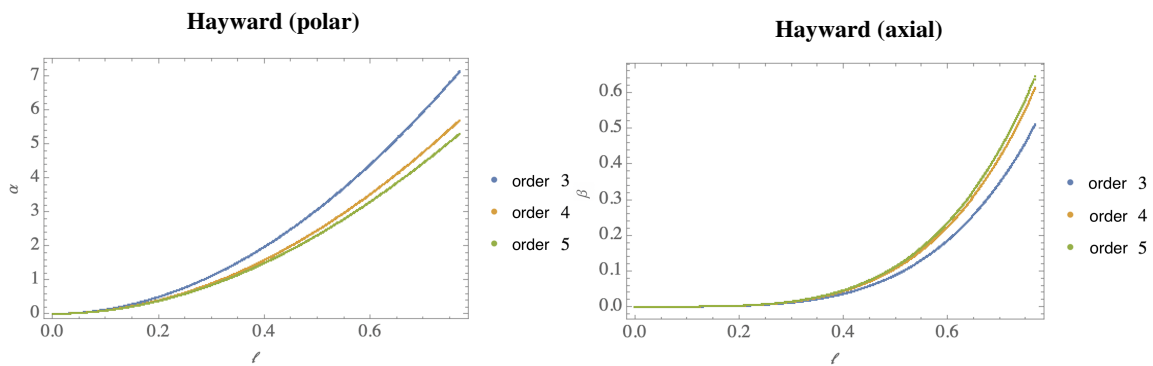

    \begin{subfigure}{0.4\textwidth}
        \centering
        \textbf{Hayward (polar)}\par\medskip
        \includegraphics[width=1.05\linewidth]{Hayward-POL-Conv.png}
    \end{subfigure}
   \hspace{2mm}
    \begin{subfigure}{0.42\textwidth}
        \centering
        \textbf{Hayward (axial)}\par\medskip
        \includegraphics[width=1.01\linewidth]{Hayward-AX-Convergence.png}
    \end{subfigure}
 \caption{{\textit{Numerical convergence.} The plots show the coefficient $\alpha$ (left) and $\beta$ (right) of the $1/r^3$ and $1/r^2$ terms, respectively,  in the asymptotic expansion of $-f(r)H_0^{20}(r)$ and $h_0^{20}(r)$ as functions of the normalized regularization parameter $\ell/\ell_\text{extremal}$ for the Hayward model. Each curve corresponds to a different truncation order $k$ in the series expansion at infinity, as indicated in the legend.}}
    \label{fig:convergence}
\end{figure}
{The truncations order k in the asymptotic expansion both at infinity and at the horizon is chosen in such a way that the related error on the reported numerical results (estimated as the difference between the result at order $k-1$ and the one at order $k$) is negligible with respect to the other uncertainties that we will now discuss.
Once the Love number curve is obtained, we fit it with a polynomial in $\ell$.  The fitting coefficients carry two main sources of uncertainty: the chosen polynomial degree and the data range used. To minimize these errors, we begin by fitting with only one power in $\ell$ over a small interval near $\ell=0$, where higher-order terms are subdominant. This allows us to identify the dominant power in $\ell$ that best represents the curve. We then progressively shrink the fitting window close to $\ell=0$ until the leading coefficient stabilizes to its final significant digits (we report only these stable decimals, so the last digit has an uncertainty of order 1). For the Bardeen and Simpson-Visser polar gravitational cases, the coefficient of the first order term in $\ell$ is small, so we also include the next term in the fit. This term has a larger coefficient and becomes comparable to the first term even at small $\ell$. For all the models, we have verified that the approximation of the curve found with the fit in a region close to $\ell=0$ remains valid across the full range up to $\ell_\text{extremal}$. Finally, by selecting several fixed values of $\ell$ and repeating the entire numerical procedure with varying $M$ for each model, we extract the dominant mass-scaling exponent through a separate fit in $M$. Consistency between this mass-fit and the $\ell$-fit (with $M=1$) confirms both the dimensional analysis and the numerical accuracy of the results.}

\bibliography{ref}

\end{document}